\documentclass[11pt]{article}

\usepackage{graphicx}
\usepackage{caption}
\usepackage[sc]{mathpazo} % Use the Palatino font
\usepackage{microtype} % Slightly tweak font spacing for aesthetics
\usepackage[T1]{fontenc}
\usepackage[utf8]{inputenc}
\usepackage{authblk}
\usepackage{comment}
\usepackage{amsmath}
\usepackage{mathtools}
\usepackage{bm}

\usepackage[english]{babel} % Language hyphenation and typographical rules
\usepackage[margin=1in]{geometry} % Document margins
\usepackage{enumitem} % Customized lists
\setlist[itemize]{noitemsep} % Make itemize lists more compact

\usepackage{url}
\usepackage{hyperref}
\usepackage{siunitx}
\usepackage{cleveref}
\usepackage{color}
\usepackage{amssymb}
\usepackage{float}
\usepackage{appendix}
\usepackage{multirow}
\usepackage{dsfont}

\usepackage{xcolor,colortbl}

\newcommand\mycolor[1]{
\ifdim #1pt>0.9 pt \cellcolor{blue!50}#1%
\else
\ifdim #1pt>0.8pt \cellcolor{blue!40}#1%
\else
\ifdim #1pt>0.7pt \cellcolor{blue!30}#1%
\else
\ifdim #1pt>0.6pt \cellcolor{blue!20}#1%
\else
\ifdim #1pt>0.5pt \cellcolor{blue!10}#1%
\else
\ifdim #1pt>0.4pt \cellcolor{red!10}#1%
\else
\ifdim #1pt>0.3pt \cellcolor{red!20}#1%
\else
\ifdim #1pt>0.2pt \cellcolor{red!30}#1%
\else
\ifdim #1pt>0.1pt \cellcolor{red!40}#1%
\else
\cellcolor{red!50}#1%
\fi\fi\fi\fi\fi\fi\fi\fi\fi
}

\DeclarePairedDelimiter\floor{\lfloor}{\rfloor}

\title{Combining Breast Cancer Risk Prediction Models}
\author[1,2]{Zoe Guan}
\author[1,2]{Theodore Huang}
\author[3]{Anne Marie McCarthy}
\author[4,5]{Kevin S. Hughes}
\author[6]{Alan Semine}
\author[1,7]{Hajime Uno}
\author[1,2]{Lorenzo Trippa}
\author[1,2]{Giovanni Parmigiani}
\author[1,2]{Danielle Braun}

\affil[1]{Department of Biostatistics, Harvard T.H. Chan School of Public Health}
\affil[2]{Department of Data Sciences, Dana-Farber Cancer Institute}
\affil[3]{Department of Biostatistics and Epidemiology, University of Pennsylvania}
\affil[4]{Division of Surgical Oncology, Massachusetts General Hospital}
\affil[5]{Department of Surgery, Harvard Medical School}
\affil[6]{Department of Radiology, Newton-Wellesley Hospital}
\affil[7]{Department of Medicine, Harvard Medical School}

\date{} % Leave empty to omit a date

\begin{document}
% Print the title
\maketitle

\begin{abstract}
Accurate risk stratification is key to reducing cancer morbidity through targeted screening and preventative interventions. Numerous breast cancer risk prediction models have been developed, but they often give predictions with conflicting clinical implications. Integrating information from different models may improve the accuracy of risk predictions, which would be valuable for both clinicians and patients. BRCAPRO and BCRAT are two widely used models based on largely complementary sets of risk factors. BRCAPRO is a Bayesian model that uses detailed family history information to estimate the probability of carrying a BRCA1/2 mutation, as well as future risk of breast and ovarian cancer, based on mutation prevalence and penetrance (age-specific probability of developing cancer given genotype). BCRAT uses a relative hazard model based on first-degree family history and non-genetic risk factors. We consider two approaches for combining BRCAPRO and BCRAT: 1) modifying the penetrance functions in BRCAPRO using relative hazard estimates from BCRAT, and 2) training an ensemble model that takes as input BRCAPRO and BCRAT predictions. We show that the combination models achieve performance gains over BRCAPRO and BCRAT in simulations and data from the Cancer Genetics Network.
\end{abstract}

\section{Introduction}

Breast cancer is the second most common cancer and the second leading cause of cancer death in women in the U.S \cite{siegel2020cancer, ACS2020}. Identifying individuals at high risk is critical for guiding decisions about risk management and prevention, including screening, genetic counseling and testing, and preventative procedures. In clinical practice, at least 24 breast cancer risk prediction models have been developed to help identify higher risk individuals \cite{cintolo2017breast}. These models estimate an individual's risk of carrying a pathogenic mutation in a breast cancer susceptibility gene and/or an individual's future risk of breast cancer, and they are based on a wide range of risk factors, methodologies, and study populations. Some models, such as the Breast Cancer Risk Assessment Tool (BCRAT) \cite{gail1989projecting, gail2007projecting, matsuno2011projecting, banegas2016projecting}, are regression-based models that use hormonal/reproductive risk factors (such as age at first live birth) and simple summaries of family history. Others, such as BRCAPRO \cite{parmigiani1998determining}, BOADICEA \cite{antoniou2004boadicea, antoniou2008boadicea, lee2019boadicea}, and IBIS \cite{tyrer2004breast}, use detailed family history information and principles of genetic inheritance. IBIS and BOADICEA \cite{lee2019boadicea} also take into account non-genetic risk factors. Different models can output risk predictions with conflicting treatment implications \cite{jacobi2009differences, ozanne2013risk}. One solution is to select a single model upon which to base intervention decisions \cite{collins2016iprevent, phillips2019accuracy}. However, identifying the best model for a given patient can be difficult, and, even if such a model is identified, other models could still contribute additional relevant information. Thus, it is important to systematically integrate information from different models to achieve more comprehensive and accurate risk assessment.

We investigate methods for combining BRCAPRO \cite{parmigiani1998determining} and BCRAT \cite{gail1989projecting, gail2007projecting, matsuno2011projecting, banegas2016projecting}, two widely used and validated \cite{spiegelman1994validation, rockhill2001validation, berry2002brcapro, amir2003evaluation, terry201910, mccarthy2019performance} breast cancer risk prediction models based on different approaches and risk factors. BRCAPRO is a family history-based model that provides carrier probabilities for breast cancer susceptibility genes BRCA1 and BRCA2 as well as future risk estimates for invasive breast cancer and for ovarian cancer. It translates family history data into risk estimates using Mendelian laws of inheritance, Bayes' rule, and literature-based estimates of mutation prevalence and penetrance (age-specific probability of developing cancer given genotype). BCRAT estimates an individual's future risk of invasive breast cancer based on a relative hazard model that includes age, hormonal and reproductive risk factors, breast biopsy history, and first-degree family history of breast cancer. The model was originally developed using case-control data from Caucasian women participating in a U.S. mammography screening program and was later updated for African-American \cite{gail2007projecting}, Asian-American \cite{matsuno2011projecting}, and Hispanic \cite{banegas2016projecting} women. 

\begin{figure}[bt]
\centering
\includegraphics[width=12cm]{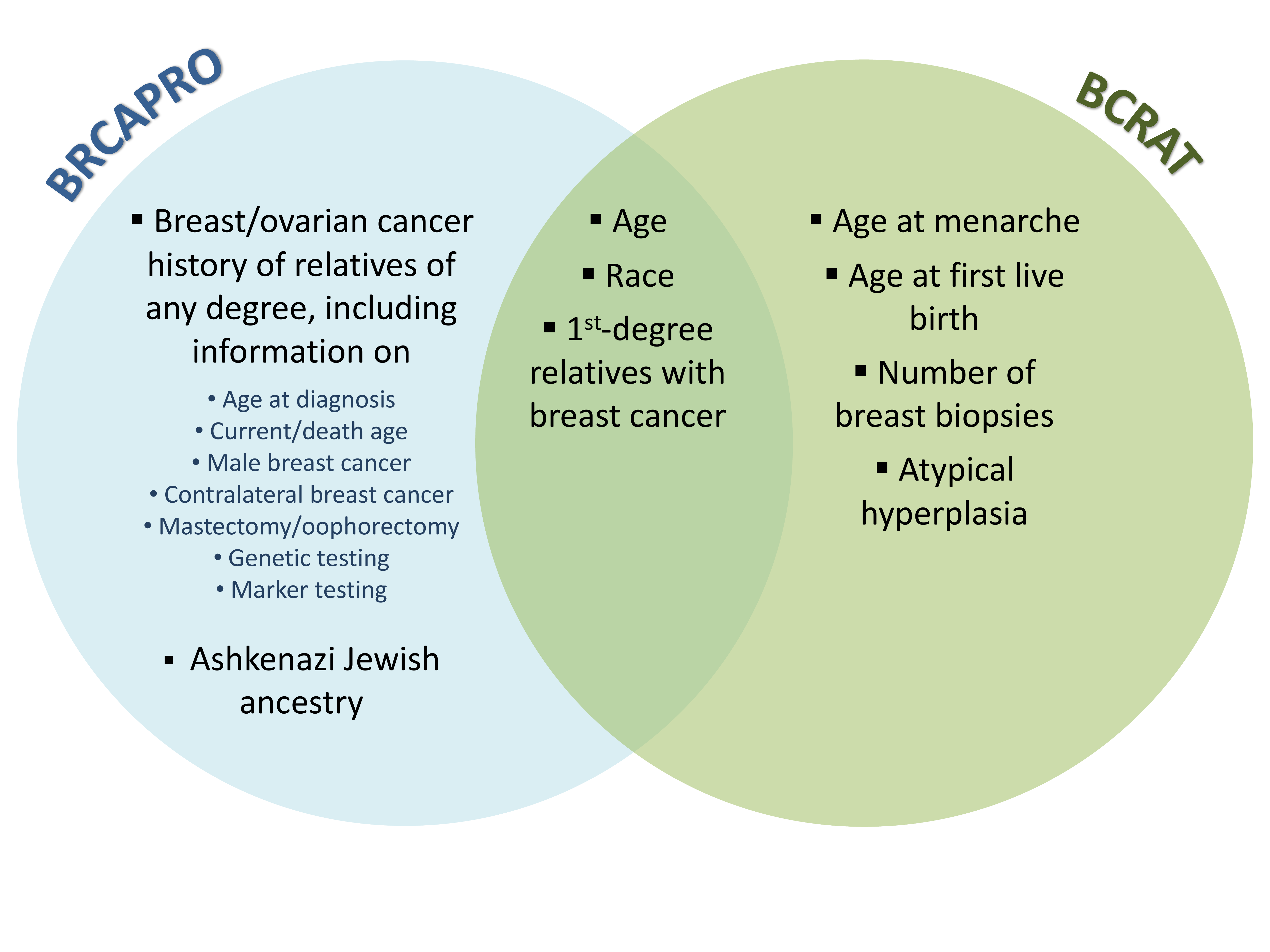}
\caption{Inputs to BRCAPRO and BCRAT.}
\label{figure:venn}
\end{figure}

Although there is some overlap in the inputs to BRCAPRO and BCRAT, the two models are largely complementary (Figure~\ref{figure:venn}). BRCAPRO uses extensive family history information while BCRAT considers only first-degree relatives. BCRAT considers several non-genetic risk factors that are not considered by BRCAPRO, including age at menarche, age at first live birth, and breast biopsies. A validation study in a large U.S. screening cohort found that 6-year risk predictions from BRCAPRO and BCRAT had a moderate correlation of 0.53 \cite{mccarthy2019performance}.  Since BRCAPRO and BCRAT embed different information, combining these models could potentially lead to accuracy gains. 

There already exist hybrid models that incorporate both detailed family history information and non-genetic risk factors: IBIS \cite{tyrer2004breast} and BOADICEA \cite{lee2019boadicea}. However, we believe it is valuable to investigate the combination of BRCAPRO and BCRAT because it will allow us to determine 1) how much predictive value non-genetic risk factors add to BRCAPRO and how much predictive value detailed family history adds to BCRAT, and 2) whether model combination can achieve competitive performance compared to developing a new hybrid model from the ground up. 

We consider two combination approaches: 1) penetrance modification and 2) training an ensemble model. The first approach involves modifying the penetrance functions in BRCAPRO to account for the effects of the BCRAT risk factors. We develop a penetrance modification model, BRCAPRO+BCRAT (M), using a relative hazard approach \cite{unpublished} that has similarities to the one used in IBIS. 

Ensemble learning consists of training multiple base models and combining their predictions. A wide variety of ensemble methods have been developed, including stacking \cite{wolpert1992stacked}, which involves training a meta-model to optimally combine predictions from the base models, bagging \cite{breiman1996bagging, breiman2001random}, which involves averaging models trained on bootstrap samples of the original data, and boosting \cite{freund1996experiments}, which involves constructing an ensemble by sequentially adding new models that are trained to correct the errors of previous ones. There is extensive literature, both empirical \cite{opitz1999popular, dietterich2000experimental} and theoretical \cite{kleinberg1990stochastic, perrone1992networks, schapire1998boosting, kuncheva2002theoretical, van2007super}, showing that ensembles can achieve performance gains over their base models, especially when the base models produce dissimilar predictions \cite{krogh1995neural, cunningham2000diversity}. In many settings, ensemble models perform well because averaging reduces variance and can expand the set of functions that can be represented by the base models \cite{dietterich2000ensemble}. Debray et al. (2014) \cite{debray2014meta} demonstrated the value of aggregating published prediction models using real and simulated data on deep venous thrombosis and traumatic brain injury. Across various scenarios, model averaging and stacking outperformed model re-calibration \cite{steyerberg2004validation, janssen2008updating} and performed as well as or better than developing a new model from scratch. Moreover, the authors noted that stacking is more efficient than model averaging since stacking has fewer unknown parameters. In the setting of breast cancer risk prediction, Ming et al. (2019) \cite{ming2019machine} showed that boosting and random forest, which is a form of bagging, were able to achieve higher discriminatory accuracy compared to BCRAT and BOADICEA. In this paper, we develop a stacked logistic regression ensemble, BRCAPRO+BCRAT (E), that takes as input predictions from BRCAPRO and BCRAT. 

We compare the performance of the combination models to the individual BRCAPRO and BCRAT models in simulations and a data application, where we use training data from the Newton-Wellesley Hospital (NWH) and validation data from the Cancer Genetics Network (CGN). In the data application, we also use IBIS as a reference for comparison to evaluate the relative performance of combining existing models versus developing a hybrid model from the ground up.

\section{Methods}

\subsection{General Notation}

Given a female proband (individual who presents for risk assessment) without a previous diagnosis of breast cancer, the goal is to predict her risk of developing invasive breast cancer within $\tau$ years based on family history $H$ (described in Section \ref{BRCAPRO}) and other risk factors $X$ (described in Section \ref{BCRAT}) while accounting for death from other causes as a competing risk. $\tau$ is a pre-specified positive integer.

Let $\tilde a$ be the proband's current age, $\tilde T_B$ the age at onset of breast cancer, $\tilde T_D$ the age at death from other causes, and $\tilde T=\min(\tilde T_B, \tilde T_D)$ the time to the first event (either breast cancer or death), with $\tilde a$, $\tilde T_B$, $\tilde T_D$, and $\tilde T$ taking on continuous values in the interval $[0, \infty)$. Let $a = \floor{\tilde a}$, $T_B = \floor{\tilde T_B}$, $T_D = \floor{\tilde T_D}$, and $T = \floor{\tilde T}$, where $\floor{\cdot}$ denotes the floor function. $a$, $T_B$, $T_D$, and $T$ are discrete versions of $\tilde a$, $\tilde T_B$, $\tilde T_D$, and $\tilde T$ observed at yearly intervals. Let $J$ denote the type of the first event, with $J=B$ if $\tilde T_B \leq \tilde T_D$ and $J=D$ otherwise. 

BRCAPRO uses a discrete model for $T$ while BCRAT and IBIS use continuous models for $\tilde T$. We will use discrete $T$ for the penetrance modification model (since it is an extension of BRCAPRO). For the ensemble model, time can be treated as either continuous or discrete. Without loss of generality, we will use continuous time notation.

\subsection{Existing Models}

\subsubsection{BRCAPRO}\label{BRCAPRO}

BRCAPRO \cite{parmigiani1998determining} estimates the probability of carrying a deleterious germline mutation in BRCA1 and BRCA2 using Bayes' rule, laws of Mendelian inheritance, mutation prevalence and penetrance, and family history. It also estimates future risk of breast and ovarian cancer based on the carrier probabilities and penetrances. 

Family history can be represented as a pedigree, or a graph where each node is a family member and edges flow from parents to offspring. Let $R$ be the number of relatives in the pedigree besides the proband and let $r=0,1,\dots,R$ index the family members, where $r=0$ corresponds to the proband. For each family member $r$, let $H_r$ be a vector containing the following information on demographics and breast/ovarian cancer history: current age or age at death, gender, race/ethnicity, relation to the proband, breast cancer status, age at breast cancer diagnosis if affected, ovarian cancer status, age at ovarian cancer diagnosis if affected, genetic testing results if available, whether the individual has had a prophylactic mastectomy, mastectomy age if applicable, whether the individual has had a prophylactic oophorectomy, and oophorectomy age if applicable. Let $H=(H_0,\dots, H_R)$. 

Additionally, let $\Gamma_r$ be family member $r$'s BRCA1/BRCA2 genotype, with $\Gamma_r=0$ if $r$ is a non-carrier, $\Gamma_r=1$ if $r$ is a carrier of a mutation in BRCA1 only, $\Gamma_r=2$ if $r$ is a carrier of a mutation in BRCA2 only, and $\Gamma_r=3$ if $r$ is a carrier of mutations in both BRCA1 and BRCA2. 

Using Bayes' rule and the assumption of conditional independence of phenotypes given genotypes (across individuals as well as cancer types), the proband's probability of having genotype $\Gamma_0$ is
\begin{equation}\label{eq:carrier}
P(\Gamma_0|H) = \frac{P(\Gamma_0) \sum_{\Gamma_1, \dots, \Gamma_R} \prod_{r=0}^R P(H_r|\Gamma_r)  P(\Gamma_1, \dots, \Gamma_R|\Gamma_0) }{\sum_{\Gamma_0} P(\Gamma_0) \sum_{\Gamma_1, \dots, \Gamma_R} \prod_{r=0}^R P(H_r|\Gamma_r)  P(\Gamma_1, \dots, \Gamma_R|\Gamma_0) }.
\end{equation}
The summation over genotypes is calculated using the Elston-Stewart peeling algorithm \cite{elston1971general} and
$P(\Gamma_1, \dots, \Gamma_R|\Gamma_0)$ is calculated based on Mendelian laws of inheritance. The prevalences $P(\Gamma_0)$ are obtained from the literature and are ethnicity-specific (in particular, different prevalences are used for Ashkenazi Jewish and non-Ashkenazi Jewish individuals). $P(H_r|\Gamma_r)$ is calculated using literature-based penetrances for breast and ovarian cancer. The penetrances are functions of age and are cancer- and sex-specific. The penetrance functions for non-carriers are based on rates from the Surveillance, Epidemiology, and End Results (SEER) program and are race-specific, while the penetrance functions for carriers are from a meta-analysis of published studies \cite{chen2020penetrance}. 

After estimating the carrier probabilities, BRCAPRO calculates future risk of breast and ovarian cancer through a weighted average of the genotype-specific risks. To simplify the notation, from here on we will omit the subscript $0$ from $\Gamma_0$ (throughout the rest of the paper, we only refer to the proband's genotype and not the genotypes of the other family members). The proband's risk of developing breast cancer between ages $a$ and $a+\tau$, conditional on having genotype $\gamma$ and not having breast cancer by age $a$, is 
\begin{flalign}
P(T \leq a+\tau, J=B|T > a, \Gamma=\gamma) %&= \sum_{t=a+1}^{a+\tau} P(T=t, J=B | T>a, \Gamma=\gamma) \\
%&= \sum_{t=a+1}^{a+\tau} P(T=t, J=B | T\geq t, T>a, \Gamma=\gamma) P(T \geq t | T>a, \Gamma=\gamma) \\
&= \sum_{t=a+1}^{a+\tau} \frac{P(T=t, J=B | T\geq t, \Gamma=\gamma) P(T \geq t|\Gamma=\gamma)}{P(T>a|\Gamma=\gamma)} \nonumber \\
&= \sum_{t=a+1}^{a+\tau} \lambda_B^\gamma(t) \prod_{u=a+1}^{t-1} \left(1-\lambda_B^\gamma(u)-\lambda_D(u) \right) \label{eq:bp_risk_g}
\end{flalign}
where $\lambda_B^\gamma(t) = P(T=t, J=B | T\geq t, \Gamma=\gamma)$ is the cause-specific hazard of breast cancer conditional on genotype $g$ and $\lambda_D(t) = P(T=t, J=D | T\geq t)$ is the cause-specific hazard of death from causes other than breast cancer. $\lambda_B^\gamma(t)$ is calculated from the female breast cancer penetrance for genotype $g$, $P(T=t, J=B | \Gamma=\gamma)$, using the recursive formula
\begin{equation}\label{eq:haz_pen}
\lambda_B^\gamma(t) = \frac{P(T=t, J=B | \Gamma=\gamma)}{\prod_{u=1}^{t-1} (1 - \lambda_B^\gamma(u)-\lambda_D(u))}, 
\end{equation}
while $\lambda_D(t)$ is estimated based on SEER mortality rates for all causes except breast cancer.

The final risk estimate is
\begin{equation}\label{eq:bp_risk}
P(T \leq a+\tau, J=B|T > a, H) = \sum_{\gamma=0}^3 P(T \leq a+\tau, J=B|T > a, \Gamma=\gamma) P(\Gamma=\gamma|H).
\end{equation}

Software for running BRCAPRO is available through the BayesMendel R package \cite{chen2004bayesmendel}. We used v2.1-6.1 (selecting the crude risk option). 

\subsubsection{BCRAT}\label{BCRAT}

BCRAT \cite{gail1989projecting, gail2007projecting, banegas2016projecting, matsuno2011projecting} estimates the relative hazard of developing breast cancer based on age (dichotomized into $<50$ and $\geq 50$) and the following risk factors: $X_1$ = age at menarche, $X_2$ = number of benign breast biopsies, $X_3$ = age at first live birth (if nulliparous, set  $X_3=25$), $X_4$ = number of female first-degree relatives with breast cancer, and $X_5$ = presence of atypical hyperplasia (0, 1, or unknown). Let $X=(X_1, \dots, X_5)$. 

The relative hazard for an individual of age $t$ with risk factors $X$ compared to an individual of age $t$ with no BCRAT risk factors (other than age) is 
\begin{flalign}
& r(t,X) = \exp (\beta_1 I[X_1 \in [12, 13]] + \beta_2 I[X_1<12]  + \nonumber \\
&\phantom{======} \beta_3 I[X_2=1] + \beta_4 I[X_2 \geq 2] + \beta_5 I[t \geq 50] I[X_2=1] + \beta_6 I[t \geq 50] I[X_2 \geq 1] + \nonumber \\
&\phantom{======} \beta_7 I[X_3 \in [20, 24]] + \beta_8 I[X_3 \in [25, 29]] + \beta_9 I[X_3 > 29] + \nonumber \\
&\phantom{======} \beta_{10} I[X_4=1] + \beta_{11} I[X_4=2] + \nonumber \\
&\phantom{======} \beta_{12} I[X_3 \in [20, 24]] I[X_4=1] + \beta_{13} I[X_3 \in [25, 29]]I[X_4=1] + \beta_{14} I[X_3 > 29]I[X_4=1] + \nonumber \\
&\phantom{======} \beta_{15} I[X_3 \in [20, 24]] I[X_4 \geq 2] + \beta_{16} I[X_3 \in [25, 29]] I[X_4 \geq 2] + \beta_{17} I[X_3 > 29] I[X_4 \geq 2] + \nonumber \\
&\phantom{======} \beta_{18} I[X_2>0] I[X_5=0] + \beta_{19} I[X_2>0] I[X_5=1]), \label{eq:rel_haz}
\end{flalign}
where $I[\cdot]$ denotes the indicator function (equal to 1 if the bracketed expression is true and 0 otherwise). The relative hazard model includes interactions between age and number of biopsies, as well as age at first live birth and number of affected relatives. The regression coefficients were estimated from U.S. case-control studies. Separate models were fit to data from white, African-American, Asian, and Hispanic women to obtain race-specific estimates. 

The risk of developing breast cancer between ages $\tilde  a$ and $\tilde a+\tau$, conditional on not having breast cancer at age $\tilde a$, is 
\begin{equation}\label{eq:bcrat_risk}
P(\tilde T \leq \tilde a+\tau, J=B| \tilde T > \tilde a, X) = \int_{\tilde a}^{\tilde a+\tau} \tilde \lambda_{B, 0}(t) r(t,X) \exp \left\{- \int_{\tilde a}^t (\tilde \lambda_{B, 0}(u)r(u,X) + \tilde \lambda_D(u)) du \right\}  dt,
\end{equation}
where $\tilde\lambda_{B, 0}(t) = \lim\limits_{dt \to 0} P(t \leq \tilde{T}<t+dt, J=B| \tilde T \geq t, X=0)/{dt}$ is the cause-specific hazard of breast cancer for those with no BCRAT risk factors and $\tilde \lambda_D(u)) = \lim\limits_{dt \to 0} P(t \leq \tilde{T}<t+dt, J=D| \tilde T \geq t)/{dt}$ is the cause-specific hazard of death from causes other than breast cancer. $\tilde \lambda_{B, 0}(t)$ is calculated from $\tilde\lambda_{B}(t) = \lim\limits_{dt \to 0} P(t \leq \tilde{T}<t+dt, J=B| \tilde T \geq t)/{dt}$, the cause-specific hazard of breast cancer in the general population, using the formula (see \cite{gail1989projecting})
\begin{equation}\label{eq:haz_ar}
\tilde\lambda_{B, 0}(t) = \tilde\lambda_{B}(t) (1-AR(t)),
\end{equation}
where, letting $P(X|t)$ be distribution of $X$ for age $t$,
\begin{equation}\label{ar}
AR(t) = 1-\frac{1}{\sum_{X} r(t,X) P(X|t)},
\end{equation}
is the population attributable risk due to $X$ for those of age $t$. $P(X|t)$ and $r(t,X)$ are both assumed to be constant for $t<50$ and for $t \geq 50$, so $AR(t)$ is as well. Race-specific estimates of $\tilde \lambda_B(t)$ and $AR(t)$ are obtained from SEER data and $\tilde \lambda_D(t)$ is estimated based on SEER mortality rates for all causes except breast cancer. In the implementation of the model, the age scale is divided into 13 intervals and $\tilde \lambda_B(t)$ and $\tilde \lambda_D(t)$ are assumed to be constant on each interval (see \cite{gail1989projecting} for more details). 

Software for running BCRAT is available through the BCRA R package (\url{https://cran.r-project.org/web/packages/BCRA/index.html}). We used version 2.1.

\subsubsection{IBIS}\label{desc_tc}

In our data application, we also use the IBIS model \cite{tyrer2004breast, brentnall2019risk} as a reference for comparison since it combines detailed family history information with non-genetic risk factors. It first calculates carrier probabilities and risk of breast cancer based on family history, then incorporates additional risk factors $Y$ (age at menarche, age at menopause, height, body mass index, age at first live birth, menopausal hormone therapy, atypical hyperplasia, lobular carcinoma in situ (LCIS), breast density, and SNPs) via a relative hazard model. The carrier probabilities are calculated using a similar approach as in BRCAPRO, but in addition to BRCA1 and BRCA2, IBIS considers a hypothetical low-penetrance susceptibility gene that acts as a surrogate for all other breast cancer susceptibility genes. The prevalence and penetrance of BRCA1 and BRCA2 are obtained from the literature and the prevalence and penetrance of the hypothetical gene are estimated using data from a Swedish population-based study. The penetrance function for non-carriers is based on rates from the Thames Cancer Registry.

IBIS calculates a weighted average of the cumulative penetrances for each genotype:
\begin{equation}\label{eq:ibis_fh}
P(\tilde T \leq t, J=B| H) = \sum_{\gamma, \theta} P(\tilde T \leq t, J=B| \Gamma, \Theta) P(\Gamma=\gamma, \Theta=\theta|H),
\end{equation}
where $\Theta$ denotes the proband's carrier status with respect to the hypothetical gene ($\Theta=0$ for non-carriers and $\Theta=1$ for carriers). In IBIS, joint carriers of BRCA1 and BRCA2 are modelled as BRCA1 carriers. 

The risk of developing breast cancer between ages $a$ and $a+\tau$, conditional on not having breast cancer at age $a$ (we use $a$ instead of $\tilde a$ here because while IBIS is based on the continuous-time framework, integer-valued ages are used in the implementation), is 
\begin{equation}\label{eq:ibis_risk}
P(\tilde T \leq a+\tau, J=B|\tilde T > a, H, Y) =  \int_{a}^{a+\tau} \tilde \lambda_{B,H}(t)s(Y)  \exp \left\{-\int_{a}^t (\tilde \lambda_{B,H}(u)s(Y) + \tilde \lambda_D(u)) du \right\} dt, 
\end{equation}
where $\tilde \lambda_{B, H}(t) = \lim\limits_{dt \to 0} P(t \leq \tilde{T}<t+dt, J=B| \tilde T \geq t, H)/{dt}$ and $s(Y)$ is a normalized version of the relative hazard of breast cancer associated with risk factors $Y$ where the normalization factor is the average relative hazard in the general population:
\begin{equation}\label{eq:ibis_haz}
 s(Y) = \phi(Y)(1-AR) =  \frac{\phi(y)}{\int \phi(Y) f(Y) dy}
 \end{equation}
where $AR$ denotes the population attributable risk due to $Y$, $\phi(Y)$ is the relative hazard associated with $Y$ (relative to the no-risk population), and $f(Y)$ is the prevalence of $Y$ in the population. $s(Y)$ is approximated by
\begin{equation}\label{eq:ibis_haz_approx}
s(Y) \approx \prod_{j=1} \frac{\phi(Y_j)}{\int \phi(Y_j)f(Y_j) dy}
 \end{equation}
using the assumption that the risk factors are independent ($j$ indexes the risk factors in $Y$).

Software for running IBIS is available at \url{http://www.ems-trials.org/riskevaluator/}. We used the command line program for version 8 and the competing mortality option.

\subsection{Model Combination Approaches}

\subsubsection{Penetrance Modification Model: BRCAPRO+BCRAT (M)}

Liu et al. \cite{unpublished} proposed to combine BRCAPRO and BCRAT by incorporating the relative hazards for BCRAT covariates into the genotype-specific hazard functions in BRCAPRO. Since BCRAT is not recommended for known carriers of pathogenic BRCA1/2 mutations (one of the patient eligibility criteria for using the risk calculator at \url{https://bcrisktool.cancer.gov/calculator.html} is the absence of a positive  BRCA1/2 test result), we propose to apply the relative hazards for the BCRAT covariates to only the non-carrier hazard function in BRCAPRO.

We extend the BCRAT relative hazard model,
\begin{equation}
\tilde \lambda_{B}(t|X) = \tilde \lambda_{B, 0}(t) r(t,X), \label{eq:bcrat_haz}
\end{equation}
where $\tilde \lambda_{B}(t|X) = \lim\limits_{dt \to 0} P(t \leq \tilde{T}<t+dt, J=B| \tilde T \geq t, X)/{dt}$, to obtain a model for non-carriers:
\begin{equation}
\tilde \lambda_{B}(t|X,\Gamma=0) = \tilde \lambda_{B}^0(t) r^0(t,X), \label{eq:bcrat_haz_g}
\end{equation}
where $\tilde \lambda_{B}(t|X, \Gamma=0) = \lim\limits_{dt \to 0} P(t \leq \tilde{T}<t+dt, J=B| \tilde T \geq t, X, \Gamma=0)/{dt}$, $\tilde \lambda_{B}^0(t) = \lim\limits_{dt \to 0} P(t \leq \tilde{T}<t+dt, J=B| \tilde T \geq t, \Gamma=0)/{dt}$, and $r^0(t,X)$ is the relative hazard of breast cancer compared to the average hazard among non-carriers (discussed in more detail below). Models \ref{eq:bcrat_haz} and \ref{eq:bcrat_haz_g} are continuous-time models. To incorporate the hazard modification into the discrete-time framework used by BRCAPRO, we consider the discrete-time analogue induced by Equation \ref{eq:bcrat_haz_g} under the setting where we observe only integer-valued $t$ (see Chapter 2.4.2 of \cite{kalbfleisch2011statistical}):
\begin{equation}\label{eq:bcrat_haz_g_discrete}
\lambda_{B}(t|X,\Gamma=0) = 1- \left(1-\lambda_{B}^0(t) \right)^{r^0(t,X)},  
\end{equation}
where $\lambda_{B}(t|X,\Gamma=0) = P(T=t, J=B | T\geq t, X, \Gamma=0)$. 

We then modify the calculation of the non-carrier risk in BRCAPRO by replacing $\lambda_{B}^0(t)$ in Equation \ref{eq:bp_risk_g} with $\lambda_{B}(t|X,\Gamma=0)$ to get
\begin{equation}\label{eq:pm_risk_g}
P(T \leq a+\tau, J=B|T > a, \Gamma=0, X) = \sum_{t=a+1}^{a+\tau} \left(  1- \left(1-\lambda_{B}^0(t) \right)^{r^0(t,X)} \right) \prod_{u=a+1}^{t-1} \left(\left(1- \lambda_{B}^0(u) \right)^{r^0(u,X)} -\lambda_D(u) \right).
\end{equation}

As in BRCAPRO, the final risk is a weighted average of the genotype-specific risks:
\begin{equation}
P(T \leq a+\tau, J=B|T > a, H, X) = \sum_{\gamma=0}^3 P(T \leq a+\tau, J=B|T > a, \Gamma=\gamma, X) P(\Gamma=\gamma|H).
\end{equation}

We refer to this model as the penetrance modification model, BRCAPRO+BCRAT (M), since the modification of the hazard function induces a modification of the corresponding penetrance function (see Equation \ref{eq:haz_pen}). This combination approach is similar to replacing the non-carrier future risk from BRCAPRO with the future risk from BCRAT, but adjusts for the slightly different baseline hazards used in BRCAPRO and BCRAT (see Figure \ref{figure:hazards} in Appendix \ref{appendix:penmod_params} for plots of the estimated general population hazard for White women in BCRAT and the estimated non-carrier hazard in BRCAPRO). 

The relative hazard approach for incorporating the BCRAT risk factors has similarities to the one used in IBIS, but IBIS averages the genotype-specific risks before incorporating non-genetic risk factors, while BRCAPRO+BCRAT (M) incorporates the BCRAT risk factors before averaging the genotype-specific risks. The advantage of the latter is that it allows for the effects of the BCRAT risk factors to differ by genotype. Differing effects by genotype have been observed for some BCRAT risk factors, such as age at menarche (see \cite{milne2016modifiers} for a review). However, in general, the effects of the BCRAT risk factors on carriers are not well-studied (only a limited number of prospective studies have been done and they had small sample sizes \cite{milne2016modifiers}), so the current version of BRCAPRO+BCRAT (M) modifies only the non-carrier hazards.

Similar to $s(Y)$ from IBIS (Equation \ref{eq:ibis_haz}), $r^0(t,X)$ is a normalized version of $r(t,X)$ where the normalization factor is the average relative hazard among non-carriers:
\begin{equation}\label{eq:pm_haz}
r^0(t,X) = r(t,X)(1-AR^0(t)) =  \frac{r(t,X)}{\sum_X r(t, X) P(X|t,\Gamma=0) dy}
 \end{equation}
where $AR^0(t)$ is the population attributable risk fraction among non-carriers. The normalization is necessary because $r(t, X)$ modifies $\tilde \lambda_{B, 0}(t)$ and is with respect to the no-risk population; in order to modify $\lambda_{B}^0(t)$, we need the relative hazard with respect to the non-carrier population. 

Due to low mutation prevalence (the prevalence of BRCA1/2 mutations in the general population has been estimated at 1/400 \cite{whittemore1997prevalence}), we approximate $AR^0(t)$ with $AR(t)$, i.e. we assume $P(X|t, \Gamma=0) \approx P(X|t)$. Therefore, BRCAPRO+BCRAT (M) takes parameters from existing models and does not need to be trained on new data (however, the parameters should be updated as new data becomes available). Though race-specific estimates of $AR(t)$ are available from BCRAT, they are based on data from the 1980s to early 2000s, so we re-estimated $AR(t)$ based on the distribution of BCRAT covariates in more recent data from the 2015 National Health Interview Survey (NHIS), which uses a cross-sectional sample of U.S. adults designed to be representative of the U.S. general population. As in IBIS (Equation \ref{eq:ibis_haz_approx}), we assumed that the risk factors are independent, except we used the joint distribution of age at first live birth and number of affected first-degree relatives because Equation \ref{eq:rel_haz} includes an interaction between these variables. The race-specific estimates from the NHIS are given in Appendix \ref{appendix:penmod_params}. 

 Since $P(\Gamma=\gamma|H)$ already accounts for family history, it may seem redundant to include the BCRAT family history variable ($X_4$, the number of affected first-degree relatives) among the penetrance-modifying risk factors. However, its inclusion could be useful because 1) there is a strong interaction between the family history variable and age at first live birth in BCRAT, and 2) the BCRAT family history variable could potentially account for residual familial risk due to non-BRCA-related factors, such as low-penetrance genes and shared environmental factors, which are not currently considered by BRCAPRO. 

\subsubsection{Ensemble Model: BRCAPRO+BCRAT (E)}

The second model combination approach involves training a stacked ensemble model \cite{wolpert1992stacked} that uses BRCAPRO and BCRAT as the (pre-trained) base models. We consider a logistic regression ensemble that predicts $\tau$-year risk of breast cancer for fixed $\tau$, as well as a time-to-event extension. These models are both special cases of the direct binomial regression model proposed by Scheike et al. (2008) \cite{scheike2008predicting} for time-to-event data with competing risks.

Let $F_B(\tau)=P(\tilde T \leq \tilde a+\tau, J=B| \tilde T > \tilde a, H, X)$. Let $F_B^{1}(\tau)$ be the $\tau$-year BRCAPRO risk prediction and $F_B^{2}(\tau)$ the $\tau$-year BCRAT risk prediction. If interest lies in risk prediction for a fixed value of $\tau$, then as in Debray et al. (2014) \cite{debray2014meta}, we can combine the $\tau$-year BRCAPRO and BCRAT predictions using the logistic regression model
\begin{equation}\label{eq:lr1}
\log \frac{F_B(\tau)}{1-F_B(\tau)} = \beta_0 + \beta_1 F_B^{1}(\tau) + \beta_2 F_B^{2}(\tau) + \beta_3 F_B^{1}(\tau) F_B^{2}(\tau).
\end{equation}
Other covariates and/or published models can also be included as inputs. We refer to Model \ref{eq:lr1} as BRCAPRO+BCRAT (E).

To predict risk for different values of $\tau$, we extend Model \ref{eq:lr1} to include interactions with $\tau$, since the optimal coefficients for combining BRCAPRO and BCRAT may vary with $\tau$:
\begin{equation}\label{eq:lr2}
\log \frac{F_B(\tau)}{1-F_B(\tau)} = \beta_0 + \beta_1 F_B^{1}(\tau) + \beta_2 F_B^{2}(\tau) + \beta_3 F_B^{1}(\tau) F_B^{2}(\tau) + \beta_4 \tau + \beta_5 \tau F_B^{1}(\tau) + \beta_6 \tau F_B^{2}(\tau) + \beta_7 \tau F_B^{1}(\tau)F_B^{2}(\tau).
\end{equation}
We refer to Model \ref{eq:lr2} as BRCAPRO + BCRAT (E2). 

Models \ref{eq:lr1} and \ref{eq:lr2} can be fit to competing risks data using generalized estimating equations with inverse probability of censoring weights (IPCW) to account for right censoring \cite{scheike2008predicting, gron2013binomial}. For Model \ref{eq:lr2}, we need to pre-specify a sequence of values of $\tau$ for applying logistic regression and use a transformed version of the training dataset where each proband contributes one observation per time point in the sequence \cite{gron2013binomial}. One choice is to use all time points at which events were observed in the training dataset \cite{gron2013binomial}. For a given $\tau$ in the pre-specified sequence, a proband's $\tau$-year event status may be unknown due to censoring. Let $\tilde C$ be the censoring time, $G(t | H, X) = P(\tilde C > t | H, X)$, $\Delta = I[\tilde T \leq \tilde C]$, and $N_B(t) = I[\tilde T \leq t, J=B]$. We assume $(\tilde T, J)$ are independent of $C$ given $(H,X)$. To account for unobserved $\tau$-year events using IPCW, we replace the outcomes $N_B(\tau)$ of those who developed cancer within $\tau$ years by $\Delta N_B(\tau) / G(\tilde T|H,X)$ \cite{scheike2008predicting} when we fit the regression model. The weighting is justified by the relationship
\[ E \left[ \frac{\Delta N_B(\tau)}{G(\tilde T | H,X)} \right] = E \left[  E \left[ \frac{\Delta N_B(\tau)}{G(\tilde T | H,X)} \Bigm\lvert T,J,H,X \right] \right] = E[N_B(\tau)|H,X] = F_B(\tau).\]
The censoring distribution $G$ can be estimated using the Kaplan-Meier method if censoring is assumed to be independent of the covariates. Otherwise, a stratified Kaplan-Meier estimator or a semiparametric regression model such as the Cox proportional hazards model could be used.

This combination approach assumes that predictions from BRCAPRO and BCRAT are not highly correlated (otherwise, multicollinearity may lead to unstable coefficient estimates \cite{su2018review}). In contrast to the penetrance modification model, the ensemble models need to be trained using prospective follow-up data. The ensemble model should ideally be trained on a dataset representative of the target population. When the training data are not representative of the target population, reweighting methods can be used to account for differences in the covariate distributions. One widely used method is importance weighting \cite{sugiyama2007covariate}, which weights each training observation by the ratio of the joint probability distributions of the covariates in the target and training populations \cite{sugiyama2007covariate}. The importance weights can be estimated using kernel mean matching \cite{huang2007correcting}, Kullback-Leibler importance estimation \cite{sugiyama2008direct}, or least squares importance fitting \cite{kanamori2009least}.

In the simulations and data application, we square root-transformed the BRCAPRO and BCRAT predictions prior to fitting/applying the ensemble models since the distributions of the predictions were highly right-skewed. We fit BRCAPRO+ BCRAT (E2) using the geepack R package (\url{https://cran.r-project.org/web/packages/geepack/index.html}), version 1.3-1, with a working autoregressive (AR-1) correlation structure. We used information from yearly time points $\tau=1, \dots, \tau^*$, where $\tau^*$ denotes the latest event time (in years from baseline) observed in the training dataset.

\subsection{Model Evaluation Metrics}

In the simulations and data application, we considered the binary outcome of being diagnosed with breast cancer within $\tau=5$ years. In the data application, we also considered the time-to-event outcome over the course of follow-up because many of the probands in the validation dataset were followed for more than 5 years, but there was a fair amount of variability in follow-up times.

We used five performance measures \cite{steyerberg2010assessing}: 1) the ratio of observed (O) to expected (E) events (where E is calculated by summing everyone's predicted probabilities), a measure of calibration (with 1 indicating perfect calibration); 2) the area under the receiver operating characteristic curve (AUC) or concordance (C) statistic, which is the probability that an individual who experiences the event has a higher score than an individual who does not and is a measure of discrimination; 3) the Brier score, which is the mean squared difference between the predicted probabilities and actual outcomes; 4) standardized net benefit (SNB) \cite{kerr2016assessing}, which is the difference between the true positive rate and a weighted false positive rate, based on a pre-specified risk threshold (the weight is the ratio of the odds of the threshold risk to the odds of the outcome); and 5) the logarithmic score \cite{good1992rational}, which is the negative log-likelihood. Calibration was assessed using both overall O/E and calibration plots of O/E by risk decile. We only calculated SNB for the binary outcome, using a 5-year risk threshold of 1.67\% (the clinical 5-year risk threshold for eligibility for chemoprevention). We report the Brier score and logarithmic score in terms of relative difference with respect to BRCAPRO since these metrics are prevalence-dependent and therefore more difficult to interpret on their original scale.

Since there was censoring in the validation data for the data application (some individuals were followed for fewer than $\tau=5$ years and were breast-cancer free when they were lost to follow-up), we used IPCW \cite{uno2007evaluating, gerds2006consistent} to calculate the O/E, AUC, Brier score, and logarithmic score for the binary outcome: individuals with observed outcomes were used to calculate the performance measures and weighted by their inverse probability of not being censored by the minimum of 1) their age at the end of the projection period, and 2) the age at which they were diagnosed with breast cancer. An individual was weighted by $1/G(a+\tau)$ if they did not develop cancer within $\tau$ years and $1/G(T_B)$ otherwise. Individuals who were censored were not directly used to calculate the performance measures, but were used to estimate the censoring distribution, $G$, via the Kaplan-Meier method. 

For the time-to-event outcome, O/E was calculated by comparing the observed to expected cases across the entire study period (for E, we predicted risk up to the end of indivudal follow-up time for each proband, so $\tau$ varied across probands). We also used time-to-event versions of the C-statistic \cite{uno2011c} and logarithmic score \cite{dawid2014theory}. The time-to-event C-statistic \cite{uno2011c} is the probability that an individual with a shorter time to event has a higher score than an individual with a longer time to event. We used a constant $\tau=10$ to calculate the C-statistic (which requires the same prediction period for everyone) since 10 years was the maximum event time observed in the data. 

We calculated 95\% bootstrap confidence intervals (CIs) for all of the performance measures except for the time-to-event C-statistic, whose 95\% CI was obtained using perturbation resampling \cite{uno2011c}.

\section{Simulations}

We compared the 5-year performance of the combination models to the individual BRCAPRO and BCRAT models in data simulated under the assumptions of the penetrance modification model. 

\subsection{Data Generation}

We first generated each proband's baseline family history, consisting of 1) the family structure, 2) dates of birth, 3) genotypes, and 4) cancer ages and death ages. 

We simulated pedigrees to mimic family structures observed in real families from the CGN dataset (the validation dataset for the data application, described in Section \ref{data}), including the number of sisters, number of brothers, and so on. We restricted the family members to first- and second-degree relatives of the proband. 

For probands, dates of birth and baseline dates for risk assessment were also sampled from the CGN dataset. For non-probands, dates of birth were generated relative to the proband's date of birth by assuming that the age difference between a parent and a child has mean 27 and standard deviation 6. We generated the birth dates of the proband's parents and children based on the proband's birth date, then the birth dates of the proband's grandparents and siblings based on the birth dates of the parents, then the birth dates of the proband's aunts and uncles based on the birth dates of the grandmothers. 

Next, we generated the BRCA genotypes for each family member. We first generated the genotypes of the grandparents using the default Ashkenazi Jewish allele frequencies for BRCA1 and BRCA2 in BRCAPRO to mimic a higher-risk population (CGN participants represent a higher-risk population than the general population since they were selected for family history of cancer). For individuals in subsequent generations, we generated genotypes according to Mendelian inheritance. 

For all individuals, we generated baseline breast and ovarian cancer phenotypes conditional on genotype. Ages of onset were sampled from \{1, 2, \dots, current age\}, with probabilities given by the genotype-specific penetrance functions from BRCAPRO, and the probability of being unaffected at baseline given by one minus the cumulative penetrance up to the current age. Probands were assumed to be alive at baseline, but we generated a death age for each non-proband from a distribution with mean 80 and standard deviation of 15. If an individual had cancer with an age of onset greater than their age at death, then the individual's cancer status was changed to unaffected. Probands with breast cancer at baseline were excluded from the analyses. 

We then generated baseline BCRAT covariates (excluding number of affected first-degree relatives and age at first live birth, which were calculated from the baseline family history), by sampling values from the distribution in the CGN. Values for different covariates were sampled independently of each other. The BCRAT covariates were used to modify the BRCAPRO non-carrier penetrance to obtain the BRCAPRO+BCRAT (M) non-carrier penetrance (Equation \ref{eq:pm_risk_g}). 

For probands who did not have breast cancer at baseline, future ages of onset were generated from the BRCAPRO+BCRAT (M) penetrances (for carriers, the BRCAPRO+BCRAT (M) penetrances are the same as in BRCAPRO), which were rescaled to be conditional on not having developed cancer by the baseline age. Cases were defined as probands who developed breast cancer within 5 years of their baseline age. The 5-year outcomes were not subject to censoring.

After excluding 4,443 probands who had breast cancer at baseline, we used 50,000 probands to train the ensemble model (similar to the size of the training set in the data application - see Section \ref{data}) and the remaining 45,557 for validation. There were 814 cases in the training set and 724 cases in the validation set.

\subsection{Results}

\begin{figure}[ht]
\centering
\includegraphics[width=0.85\columnwidth]{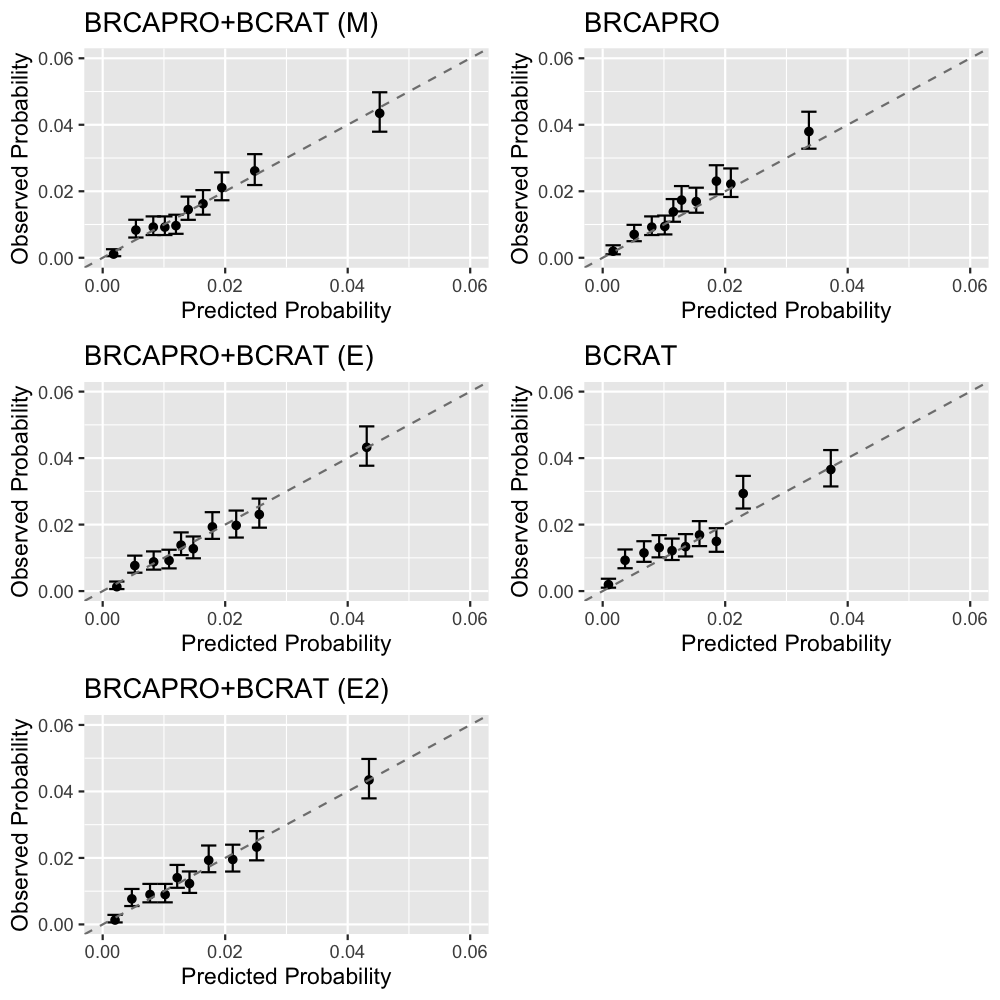}
\caption{Calibration plots by decile of risk for 5-year predictions in a simulated dataset with 45,557 probands (724 cases).}
\label{figure:calplots_sim}
\end{figure}

The performance measures are given in Table \ref{table:table_sim} and calibration plots are given in Figure \ref{figure:calplots_sim}. BRCAPRO+BRCAT (M), the true model, had the best performance, but the ensemble models were able to achieve similar performance to the true model and performance gains over BRCAPRO and BCRAT. The combination models were well-calibrated overall, with O/E=1.01 (95\% CI 0.94-1.08) for BRCAPRO+BCRAT (M), O/E=0.98 (95\% CI 0.91-1.04) for BRCAPRO+BCRAT (E), and O/E=1.00 (95\% CI 0.93-1.07) for BRCAPRO+BCRAT (E2), while BRCAPRO and BCRAT underpredicted the number of cases, with O/E=1.14 (95\% CI 1.07-1.22) for BRCAPRO and O/E=1.13 (95\% CI 1.05-1.20) for BCRAT. The combination models were well-calibrated in each decile of risk (Figure \ref{figure:calplots_sim}), while BRCAPRO and BCRAT were more prone to underpredicting risk in certain deciles. The combination models had slightly higher AUCs than BRCAPRO and BCRAT: 0.69 (95\% CI 0.67-0.71 for BRCAPRO+BCRAT (M), 0.68 (95\% CI 0.67-0.70) for BRCAPRO+BCRAT (E) and BRCAPRO+BCRAT (E2), 0.67 (95\% CI 0.65-0.69) for BRCAPRO, and 0.66 (95\% CI 0.64-0.68) for BRCAT. Also, the combination models performed better than BRCAPRO and BCRAT with respect to the Brier score, logarithmic score, and SNB. Across 1000 bootstrap replicates of the validation dataset, all of the combination models outperformed BRCAPRO and BCRAT with respect to all performance measures in more than 95\% of the replicates. BRCAPRO+BCRAT (M) outperformed BRCAPRO+BCRAT (E) and BRCAPRO+BCRAT (E2) more than 99\% of the time with respect to AUC, Brier score, and logarithmic score.

% latex table generated in R 3.5.2 by xtable 1.8-3 package
% Fri Jul 31 08:10:00 2020
\begin{table}[!htbp]
\centering
\begingroup\footnotesize
\begin{tabular}{llllll}
  \hline
 & O/E & AUC & SNB & $\Delta$BS & $\Delta$LS \\ 
  \hline \textbf{Performance Metrics} & \\ 
B+B (M) & 1.01 (0.94, 1.08) & 0.69 (0.67, 0.71) & 0.26 (0.21, 0.30) & 0.25 (0.09, 0.41) & 1.12 (0.60, 1.66) \\ 
  B+B (E) & 0.98 (0.91, 1.04) & 0.68 (0.67, 0.70) & 0.25 (0.20, 0.29) & 0.12 (-0.02, 0.25) & 0.63 (0.18, 1.05) \\ 
  B+B (E2) & 1.00 (0.93, 1.07) & 0.68 (0.67, 0.70) & 0.24 (0.19, 0.28) & 0.13 (-0.03, 0.27) & 0.61 (0.17, 1.06) \\ 
  BRCAPRO & 1.15 (1.07, 1.23) & 0.67 (0.65, 0.69) & 0.21 (0.17, 0.25) & 0.00 (0.00, 0.00) & 0.00 (0.00, 0.00) \\ 
  BCRAT & 1.14 (1.05, 1.21) & 0.66 (0.64, 0.68) & 0.20 (0.15, 0.24) & -0.21 (-0.46, 0.06) & -1.15 (-2.10, -0.16) \\ 
  B+B(M)>B+B(E) & \mycolor{0.570} & \mycolor{0.994} & \mycolor{0.725} & \mycolor{0.999} & \mycolor{1.000} \\ 
   \hline \multicolumn{3}{l}{\textbf{Comparisons Across Bootstrap Replicates}} & & & \\ 
B+B(M)>B+B(E2) & \mycolor{0.422} & \mycolor{0.994} & \mycolor{0.886} & \mycolor{0.995} & \mycolor{1.000} \\ 
  B+B(M)>BRCAPRO & \mycolor{0.978} & \mycolor{0.999} & \mycolor{0.998} & \mycolor{0.998} & \mycolor{1.000} \\ 
  B+B(M)>BCRAT & \mycolor{0.962} & \mycolor{1.000} & \mycolor{1.000} & \mycolor{1.000} & \mycolor{1.000} \\ 
  B+B(E)>B+B(E2) & \mycolor{0.401} & \mycolor{0.625} & \mycolor{0.920} & \mycolor{0.235} & \mycolor{0.653} \\ 
  B+B(E)>BRCAPRO & \mycolor{0.944} & \mycolor{0.999} & \mycolor{1.000} & \mycolor{0.961} & \mycolor{0.998} \\ 
  B+B(E)>BCRAT & \mycolor{0.924} & \mycolor{1.000} & \mycolor{0.999} & \mycolor{0.994} & \mycolor{1.000} \\ 
  B+B(E2)>BRCAPRO & \mycolor{0.974} & \mycolor{0.999} & \mycolor{0.991} & \mycolor{0.957} & \mycolor{0.996} \\ 
  B+B(E2)>BCRAT & \mycolor{0.952} & \mycolor{1.000} & \mycolor{0.998} & \mycolor{0.995} & \mycolor{1.000} \\ 
   \hline
\end{tabular}
\endgroup
\caption{5-year performance in a simulated dataset with 45,557 probands (717 cases). B+B: BRCAPRO+BCRAT. $\Delta$BS: \% relative improvement in Brier Score compared to BRCAPRO. $\Delta$LS: \% relative improvement in logarithmic score compared to BRCAPRO. The ``Comparisons Across Bootstrap Replicates" section shows pairwise comparisons involving the combination models across 1000 bootstrap replicates of the validation dataset; the row for $A>B$ shows the proportion of bootstrap replicates where model A outperformed model B with respect to each metric. Proportions $>0.5$ are highlighted in blue (with darker shades of blue for higher proportions) and proportions $\leq 0.5$ are highlighted in red (with darker shades of red for lower proportions).} 
\label{table:table_sim}
\end{table}

\section{Data Application}

We trained ensemble models \ref{eq:lr1} for $\tau=5$ and \ref{eq:lr2} for time-to-event outcome using data from the Newton-Wellesley Hospital (NWH) and validated them, along with BRCAPRO+BCRAT (M), BRCAPRO, BCRAT, and IBIS, on data from the CGN. We assessed performance based on both the binary outcome of breast cancer diagnosis within $\tau=5$ years and the time-to-event outcome. We also looked at performance stratified by family history, with strata defined based on the NCCN criteria for further genetic risk evaluation \cite{nccn}.

In the analyses, we excluded women with invasive breast cancer/ductal carcinoma in situ/lobular carcinoma in situ/bilateral mastectomy/bilateral oophorectomy prior to baseline, women who tested positive for BRCA1/2 prior to baseline (BCRAT requirement), women $<20$ years old at baseline (BCRAT requirement), and women with projection age $>85$ years old (IBIS requirement).

\subsection{Datasets}\label{data}

The characteristics of the training and validation datasets are summarized in Table \ref{table:datasets}.

\subsubsection{NWH}
After applying the exclusion criteria, the training cohort consisted of 37,881 women who visited the breast imaging department of the NWH in Newton, Massachusetts for screening or diagnostic imaging from February 2007 through December 2009. During the initial (baseline) visit, information was collected on personal and family history of cancer, reproductive history, sociodemographic factors, and lifestyle factors. Family history was limited to relatives with cancer. Breast cancer diagnoses through 2015 were determined from the Massachusetts State Cancer Registry, Partners Hospital Cancer Registries, and patient self-reporting. The median age of the probands was 49, with an inter-quartile range (IQR) of 43-58. 30,758 (81.2\%) of the probands were White. 5,684 (15.0\%) had at least one affected first- or second-degree relative. The median follow-up time was 6.7 years (IQR 6.3-7.2). All probands were followed for at least 6 years, so we did not need to use IPCW to fit BRCAPRO+BCRAT (E) for $\tau=5$ years. To fit BRCAPRO+BCRAT (E2), we used the Kaplan-Meier method to estimate the censoring distribution for IPCW. There were 495 probands (1.3\%) who developed breast cancer within 5 years of baseline and 714 probands (1.9\%) who developed breast cancer over the course of follow-up.

Since the NWH cohort represents a general screening population while the CGN validation cohort (described below) represents a higher-risk population enriched for family history of cancer, we applied importance weights to the training data based on the distributions of the BCRAT covariates, 5-year BCRAT predictions, and 5-year BRCAPRO predictions when we fit the ensemble models. The weights were estimated using least squares importance fitting via the densratio R package \url{https://cran.r-project.org/web/packages/densratio/index.html}.  

\subsubsection{CGN}
The validation cohort consists of 7,314 women who enrolled in the CGN, a national research network consisting of 15 academic medical centers that was established for the purpose of studying inherited predisposition to cancer. Enrollment began in 1999 and ended in 2010. One of the criteria for enrollment was a personal and/or family history of cancer. Participants provided information on personal and family history of cancer, sociodemographic factors and lifestyle factors through an initial (baseline) phone interview and annual follow-up updates. From 2009 onward, information was also collected on reproductive history, cancer treatments, cancer screening results, and genetic testing results. 

The median age of the probands was 47 (IQR 38-57); 6,104 (83.5\%) of the probands were White. 3,143 (42.9\%) had at least one female first-degree relative with breast cancer. The median follow-up time was 7.3 years (IQR 6.0-8.3) and 934 (12.8\%) probands were censored within 5 years of baseline without being diagnosed with breast cancer. 159 (2.2\%) probands developed breast cancer during follow-up, with 112 of the diagnoses occurring within 5 years of baseline. Demographic characteristics stratified by center are given in Table \ref{table:cgn_centers} in Appendix \ref{appendix:cgn_supp}. Since follow-up times and breast cancer incidence rates varied by center, we estimated the censoring distribution by fitting a Kaplan-Meier curve for each center separately.

Information on some risk factors was missing or incomplete. We did not have information on atypical hyperplasia (used in BCRAT and IBIS), breast density (used in IBIS), polygenic risk scores (used in IBIS, or hormone replacement therapy (used in IBIS). Participants were asked whether they had ever had a benign breast biopsy but were not asked about the number of biopsies (categorized as 0, 1, or $\geq 2$ in BCRAT). Since participants were asked about reproductive history starting in 2009, 4,157 (56.8\%) were missing age at menarche (used in BCRAT and IBIS). Ashkenazi Jewish status (used in BRCAPRO and IBIS) was not available for the UWASH center. We coded the missing variables according to the specifications of the software for each model. Number of breast biopsies was coded as 1 for participants who indicated that they had previously had a biopsy.

\begin{table}[ht]
\centering
\begin{tabular}{llll}
  \hline
Variable & Category & CGN & NWH \\ 
  \hline
N &  & 7314 & 37881 \\ 
  Age (median [IQR]) &  & 47 [38, 57] & 49 [43, 58] \\ 
  Race (\%) & White & 6104 (83.5) & 30758 (81.2) \\ 
   & Black & 257 (3.5) & 479 (1.3) \\ 
   & Hispanic & 694 (9.5) & 548 (1.4) \\ 
   & Asian & 160 (2.2) & 1228 (3.2) \\ 
   & Native American & 29 (0.4) & 25 (0.1) \\ 
   & Unknown & 70 (1.0) & 4843 (12.8) \\ 
  Affected 1st-degree Relatives (\%) & 0 & 4171 (57.0) & 32197 (85.0) \\ 
   & 1 & 2496 (34.1) & 5277 (13.9) \\ 
   & 2+ & 647 (8.8) & 407 (1.1) \\ 
  Follow-up Time (median [IQR]) &  & 7.3 [6.0, 8.3] & 6.7 [6.3, 7.2] \\ 
  Censored $<$5 Years (\%) &  & 934 (12.8) & 0 (0.0) \\ 
  Cases (\%) &  & 159 (2.2) & 714 (1.9) \\ 
  5-year Cases (\%) &  & 112 (1.5) & 495 (1.3) \\ 
   \hline
\end{tabular}
\caption{Cohort characteristics.} 
\label{table:datasets}
\end{table}

\subsection{Results}

\subsubsection{5-year Binary Outcome}

\begin{figure}[ht]
\centering
\includegraphics[width=0.85\columnwidth]{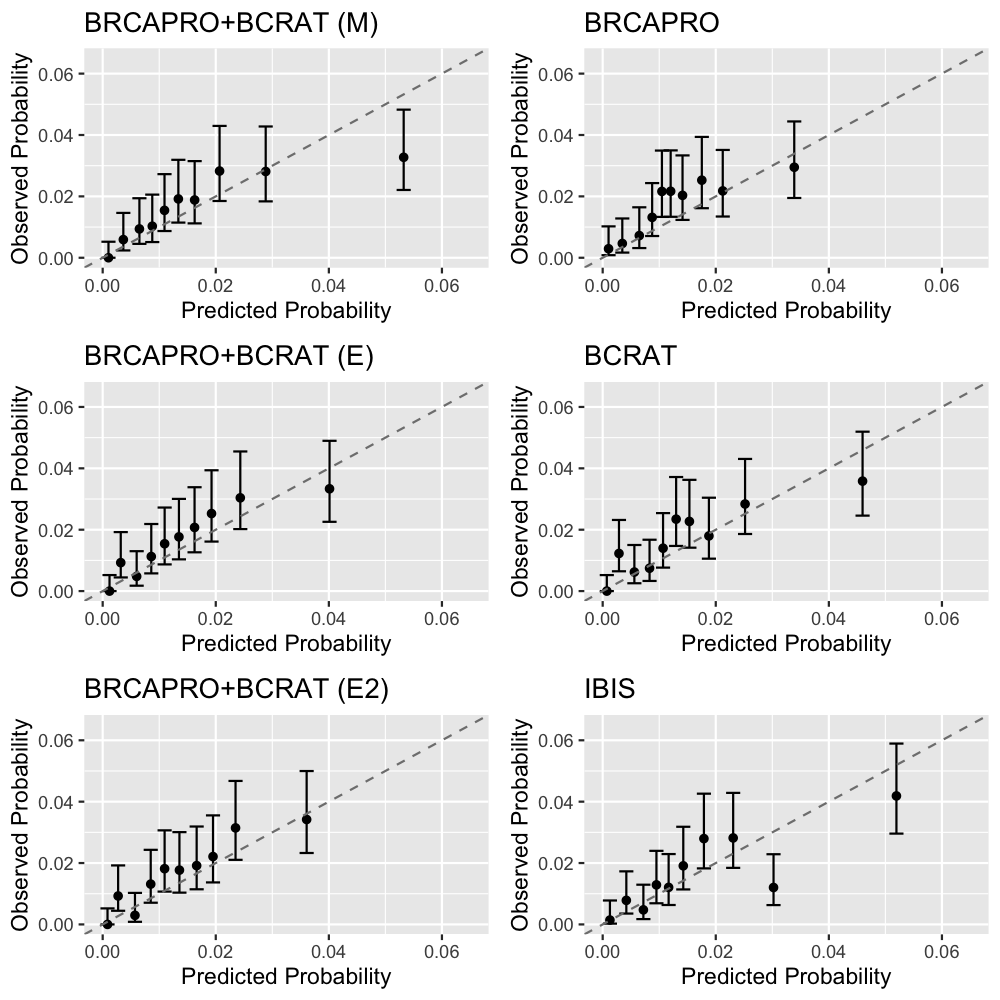}
\caption{Calibration plots.}
\label{figure:calplots}
\end{figure}

The performance measures based on the 5-year outcome are shown in Tables \ref{table:cgn5} (overall performance) and \ref{table:cgn5_fh} (performance stratified by family history). Calibration plots are shown in Figure \ref{figure:calplots} and scatter plots, density plots, and correlations are shown in Figure \ref{figure:scatterplots}. The weights from the ensemble models are provided in Appendix \ref{appendix:weights}. 

Predictions from BRCAPRO+BCRAT (M) were highly correlated with predictions from each of the other models in the entire cohort (Figure \ref{figure:scatterplots}), with Pearson correlation coefficients ranging from $\rho=0.78$ with BRCAPRO to $\rho=0.93$ with BRCAPRO+BCRAT (E). BRCAPRO+BCRAT (E), which assigned a higher weight to BCRAT than to BRCAPRO (see Appendix \ref{appendix:weights}), was very highly correlated with BCRAT ($\rho=0.93$) and moderately correlated with BRCAPRO ($\rho=0.67$), while BRCAPRO+BCRAT (E2) assigned similar weights to BRCAPRO and BCRAT and was highly correlated with both models ($\rho=0.80$ and $\rho=0.75$).

BRCAPRO+BCRAT (M) (O/E=1.03, 95\% CI 0.85-1.23) and IBIS (O/E=0.98, 95\% CI 0.81-1.17) well-calibrated overall while BCRAT (O/E=1.15, 95\% CI 0.95-1.37), BRCAPRO+BCRAT (E) (O/E=1.17, 95\% CI 0.97-1.40), BRCAPRO+BCRAT (E2) (O/E=1.22, 95\% CI 1.00-1.45), and BRCAPRO (O/E=1.30, 95\% CI 1.07-1.55) underestimated risk. BRCAPRO+BCRAT (M) overestimated risk in the top decile of risk and IBIS overestimated risk in the second highest decile, while BRCAPRO and BCRAT underestimated risk in several deciles (Figure \ref{figure:calplots}). The AUCs were 0.68 (95\% CI 0.63-0.72) for the combination models, 0.67 for IBIS (95\% CI 0.62-0.71), 0.66 for BCRAT (95\% CI 0.62-0.71), and 0.65 (95\% CI 0.61-0.70) for BRCAPRO. IBIS had the highest SNB (SNB=0.28, 95\% CI 0.16-0.39), followed by BRCAPRO+BCRAT (M) (SNB=0.24, 95\% CI 0.13-0.35) and the ensemble models (SNB=0.23, 95\% CI 0.11-0.33 for (E) and 0.11-0.34 for (E2)). All models performed similarly with respect to the Brier score and logarithmic score. Across 1000 bootstrap replicates, BRCAPRO+BCRAT (M) outperformed BRCAPRO and BCRAT with respect to all performance measures except the Brier score in the majority of the replicates. In particular, BRCAPRO+BCRAT (M) had a higher SNB than BRCAPRO /BCRAT 95\% of the time or more. Both ensemble models outperformed BRCAPRO and BCRAT with respect to all metrics except O/E in the majority of the replicates. Also, the combination models outperformed IBIS with respect to AUC, Brier score, and logarithmic score in the majority of the replicates, but IBIS had better calibration and SNB in most replicates.

In probands who met the NCCN criteria for further genetic risk evaluation (Table \ref{table:cgn5_fh}), the combination models and IBIS had higher AUCs and SNBs than BRCAPRO and BCRAT.  BRCAPRO+BCRAT (M) and IBIS overestimated risk while all other models except BCRAT underestimated risk. In probands who did not meet the NCCN criteria, all models underestimated risk. BRCAPRO had a slightly lower AUC than the other models and IBIS had the highest SNB.

% latex table generated in R 3.5.2 by xtable 1.8-3 package
% Fri Jul  3 09:04:21 2020
\begin{table}[!htbp]
\centering
\begingroup\footnotesize
\begin{tabular}{llllll}
  \hline
 & O/E & AUC & SNB & $\Delta$BS & $\Delta$LS \\ 
  \hline \textbf{Performance Metrics} & \\ 
B+B (M) & 1.03 (0.85, 1.23) & 0.68 (0.63, 0.72) & 0.24 (0.13, 0.35) & 0.21 (-0.43, 0.84) & 1.61 (-0.54, 3.64) \\ 
  B+B (E) & 1.17 (0.97, 1.40) & 0.68 (0.63, 0.72) & 0.23 (0.11, 0.33) & 0.38 (-0.09, 0.90) & 1.74 (-0.10, 3.52) \\ 
  B+B (E2) & 1.22 (1.00, 1.45) & 0.68 (0.63, 0.72) & 0.23 (0.11, 0.34) & 0.32 (0.00, 0.66) & 1.58 (0.17, 3.04) \\ 
  BRCAPRO & 1.30 (1.07, 1.55) & 0.65 (0.61, 0.70) & 0.17 (0.06, 0.27) & 0.00 (0.00, 0.00) & 0.00 (0.00, 0.00) \\ 
  BCRAT & 1.15 (0.95, 1.37) & 0.66 (0.62, 0.71) & 0.18 (0.07, 0.29) & 0.22 (-0.53, 0.97) & 0.88 (-2.05, 3.47) \\ 
  IBIS & 0.98 (0.81, 1.17) & 0.67 (0.62, 0.71) & 0.28 (0.16, 0.39) & -0.05 (-0.69, 0.51) & 0.89 (-1.56, 2.95) \\ 
   \hline \multicolumn{3}{l}{\textbf{Comparisons Across Bootstrap Replicates}} & & & \\ 
B+B(M)>B+B(E) & \mycolor{0.839} & \mycolor{0.418} & \mycolor{0.680} & \mycolor{0.203} & \mycolor{0.399} \\ 
  B+B(M)>B+B(E2) & \mycolor{0.892} & \mycolor{0.439} & \mycolor{0.609} & \mycolor{0.327} & \mycolor{0.508} \\ 
  B+B(M)>BRCAPRO & \mycolor{0.950} & \mycolor{0.931} & \mycolor{0.950} & \mycolor{0.757} & \mycolor{0.932} \\ 
  B+B(M)>BCRAT & \mycolor{0.807} & \mycolor{0.760} & \mycolor{0.980} & \mycolor{0.456} & \mycolor{0.778} \\ 
  B+B(M)>IBIS & \mycolor{0.460} & \mycolor{0.745} & \mycolor{0.191} & \mycolor{0.825} & \mycolor{0.789} \\ 
  B+B(E)>B+B(E2) & \mycolor{0.969} & \mycolor{0.578} & \mycolor{0.408} & \mycolor{0.703} & \mycolor{0.708} \\ 
  B+B(E)>BRCAPRO & \mycolor{0.988} & \mycolor{0.960} & \mycolor{0.914} & \mycolor{0.935} & \mycolor{0.969} \\ 
  B+B(E)>BCRAT & \mycolor{0.055} & \mycolor{0.849} & \mycolor{0.867} & \mycolor{0.871} & \mycolor{0.906} \\ 
  B+B(E)>IBIS & \mycolor{0.223} & \mycolor{0.749} & \mycolor{0.150} & \mycolor{0.909} & \mycolor{0.813} \\ 
  B+B(E2)>BRCAPRO & \mycolor{0.995} & \mycolor{0.979} & \mycolor{0.940} & \mycolor{0.976} & \mycolor{0.986} \\ 
  B+B(E2)>BCRAT & \mycolor{0.038} & \mycolor{0.785} & \mycolor{0.870} & \mycolor{0.659} & \mycolor{0.756} \\ 
  B+B(E2)>IBIS & \mycolor{0.163} & \mycolor{0.735} & \mycolor{0.190} & \mycolor{0.893} & \mycolor{0.768} \\ 
   \hline
\end{tabular}
\endgroup
\caption{5-year performance in the entire CGN cohort. B+B: BRCAPRO+BCRAT. $\Delta$BS: \% relative improvement in Brier Score compared to BRCAPRO. $\Delta$LS: \% relative improvement in logarithmic score compared to BRCAPRO. The ``Comparisons Across Bootstrap Replicates" section shows pairwise comparisons involving the combination models across 1000 bootstrap replicates of the validation dataset; the row for $A>B$ shows the proportion of bootstrap replicates where model A outperformed model B with respect to each metric. Proportions $>0.5$ are highlighted in blue (with darker shades of blue for higher proportions) and proportions $\leq 0.5$ are highlighted in red (with darker shades of red for lower proportions).} 
\label{table:cgn5}
\end{table}

% latex table generated in R 3.5.2 by xtable 1.8-3 package
% Fri Jul  3 09:44:21 2020
\begin{table}[!htbp]
\centering
\begingroup\scriptsize
\begin{tabular}{llllll}
  \hline
 & O/E & AUC & SNB & $\Delta$BS & $\Delta$LS \\ 
  \hline \multicolumn{3}{l}{\underline{\textbf{Strong Family History (34 cases)}}} & & & \\
\textbf{Performance Metrics} & & & & & \\ 
B+B (M) & 0.81 (0.55, 1.09) & 0.71 (0.63, 0.79) & 0.44 (0.20, 0.59) & 0.75 (-1.41, 2.29) & 3.62 (-3.05, 7.97) \\ 
  B+B (E) & 1.07 (0.73, 1.44) & 0.71 (0.62, 0.79) & 0.41 (0.17, 0.57) & 1.19 (-0.21, 2.48) & 4.15 (-0.75, 8.42) \\ 
  B+B (E2) & 1.16 (0.78, 1.55) & 0.71 (0.63, 0.79) & 0.41 (0.17, 0.57) & 1.03 (0.12, 1.88) & 3.86 (0.03, 7.00) \\ 
  BRCAPRO & 1.32 (0.88, 1.79) & 0.66 (0.58, 0.74) & 0.30 (0.09, 0.47) & 0.00 (0.00, 0.00) & 0.00 (0.00, 0.00) \\ 
  BCRAT & 1.03 (0.70, 1.38) & 0.66 (0.57, 0.76) & 0.31 (0.06, 0.47) & 0.77 (-1.42, 2.70) & 1.55 (-6.27, 8.10) \\ 
  IBIS & 0.74 (0.50, 0.99) & 0.69 (0.61, 0.77) & 0.41 (0.14, 0.57) & -0.15 (-2.38, 1.33) & 1.12 (-5.97, 5.56) \\ 
   \hline \multicolumn{3}{l}{\textbf{Comparisons Across Bootstrap Replicates}} & & & \\ 
B+B(M)>B+B(E) & \mycolor{0.344} & \mycolor{0.607} & \mycolor{0.645} & \mycolor{0.183} & \mycolor{0.321} \\ 
  B+B(M)>B+B(E2) & \mycolor{0.445} & \mycolor{0.513} & \mycolor{0.679} & \mycolor{0.325} & \mycolor{0.434} \\ 
  B+B(M)>BRCAPRO & \mycolor{0.609} & \mycolor{0.939} & \mycolor{0.905} & \mycolor{0.792} & \mycolor{0.885} \\ 
  B+B(M)>BCRAT & \mycolor{0.284} & \mycolor{0.908} & \mycolor{0.952} & \mycolor{0.462} & \mycolor{0.766} \\ 
  B+B(M)>IBIS & \mycolor{0.954} & \mycolor{0.692} & \mycolor{0.707} & \mycolor{0.869} & \mycolor{0.909} \\ 
  B+B(E)>B+B(E2) & \mycolor{0.707} & \mycolor{0.436} & \mycolor{0.790} & \mycolor{0.666} & \mycolor{0.616} \\ 
  B+B(E)>BRCAPRO & \mycolor{0.824} & \mycolor{0.851} & \mycolor{0.896} & \mycolor{0.949} & \mycolor{0.952} \\ 
  B+B(E)>BCRAT & \mycolor{0.416} & \mycolor{0.953} & \mycolor{0.866} & \mycolor{0.837} & \mycolor{0.941} \\ 
  B+B(E)>IBIS & \mycolor{0.747} & \mycolor{0.639} & \mycolor{0.537} & \mycolor{0.916} & \mycolor{0.889} \\ 
  B+B(E2)>BRCAPRO & \mycolor{0.874} & \mycolor{0.902} & \mycolor{0.890} & \mycolor{0.987} & \mycolor{0.976} \\ 
  B+B(E2)>BCRAT & \mycolor{0.334} & \mycolor{0.885} & \mycolor{0.863} & \mycolor{0.634} & \mycolor{0.801} \\ 
  B+B(E2)>IBIS & \mycolor{0.641} & \mycolor{0.669} & \mycolor{0.515} & \mycolor{0.893} & \mycolor{0.871} \\ 
   \hline \multicolumn{3}{l}{\underline{\textbf{Less Family History (78 cases)}}} & & & \\
\textbf{Performance Metrics} & & & & & \\ 
 \hline
B+B (M) & 1.16 (0.91, 1.42) & 0.65 (0.60, 0.71) & 0.16 (0.02, 0.28) & -0.02 (-0.48, 0.37) & 0.80 (-1.15, 2.55) \\ 
  B+B (E) & 1.21 (0.95, 1.49) & 0.66 (0.60, 0.71) & 0.14 (-0.01, 0.27) & 0.03 (-0.22, 0.26) & 0.80 (-0.64, 2.18) \\ 
  B+B (E2) & 1.24 (0.97, 1.51) & 0.66 (0.60, 0.71) & 0.15 (-0.01, 0.28) & 0.02 (-0.18, 0.21) & 0.70 (-0.48, 1.90) \\ 
  BRCAPRO & 1.28 (1.00, 1.57) & 0.63 (0.58, 0.68) & 0.10 (-0.04, 0.22) & 0.00 (0.00, 0.00) & 0.00 (0.00, 0.00) \\ 
  BCRAT & 1.20 (0.94, 1.47) & 0.66 (0.60, 0.72) & 0.12 (-0.02, 0.25) & -0.01 (-0.43, 0.36) & 0.68 (-1.68, 3.02) \\ 
  IBIS & 1.14 (0.89, 1.39) & 0.66 (0.60, 0.71) & 0.22 (0.07, 0.35) & -0.01 (-0.43, 0.38) & 0.81 (-1.40, 2.95) \\ 
   \hline \multicolumn{3}{l}{\textbf{Comparisons Across Bootstrap Replicates}} & & & \\ 
B+B(M)>B+B(E) & \mycolor{0.918} & \mycolor{0.294} & \mycolor{0.646} & \mycolor{0.336} & \mycolor{0.487} \\ 
  B+B(M)>B+B(E2) & \mycolor{0.927} & \mycolor{0.428} & \mycolor{0.555} & \mycolor{0.361} & \mycolor{0.550} \\ 
  B+B(M)>BRCAPRO & \mycolor{0.949} & \mycolor{0.860} & \mycolor{0.815} & \mycolor{0.446} & \mycolor{0.789} \\ 
  B+B(M)>BCRAT & \mycolor{0.912} & \mycolor{0.248} & \mycolor{0.898} & \mycolor{0.429} & \mycolor{0.557} \\ 
  B+B(M)>IBIS & \mycolor{0.132} & \mycolor{0.412} & \mycolor{0.122} & \mycolor{0.434} & \mycolor{0.483} \\ 
  B+B(E)>B+B(E2) & \mycolor{0.948} & \mycolor{0.814} & \mycolor{0.389} & \mycolor{0.539} & \mycolor{0.710} \\ 
  B+B(E)>BRCAPRO & \mycolor{0.960} & \mycolor{0.945} & \mycolor{0.791} & \mycolor{0.607} & \mycolor{0.867} \\ 
  B+B(E)>BCRAT & \mycolor{0.078} & \mycolor{0.383} & \mycolor{0.678} & \mycolor{0.714} & \mycolor{0.579} \\ 
  B+B(E)>IBIS & \mycolor{0.094} & \mycolor{0.510} & \mycolor{0.113} & \mycolor{0.588} & \mycolor{0.494} \\ 
  B+B(E2)>BRCAPRO & \mycolor{0.965} & \mycolor{0.956} & \mycolor{0.850} & \mycolor{0.582} & \mycolor{0.879} \\ 
  B+B(E2)>BCRAT & \mycolor{0.058} & \mycolor{0.322} & \mycolor{0.710} & \mycolor{0.613} & \mycolor{0.497} \\ 
  B+B(E2)>IBIS & \mycolor{0.080} & \mycolor{0.459} & \mycolor{0.155} & \mycolor{0.570} & \mycolor{0.447} \\ 
  \end{tabular}
\endgroup
\caption{5-year performance in the CGN cohort, stratified by family history (whether or not the proband met the NCCN criteria for further genetic risk evaluation \cite{nccn}; in applying the criteria, we only used information on breast and ovarian cancer diagnoses in relatives). B+B: BRCAPRO+BCRAT. $\Delta$BS: \% relative improvement in Brier Score compared to BRCAPRO. The ``Comparisons Across Bootstrap Replicates" section shows pairwise comparisons involving the combination models across 1000 bootstrap replicates of the validation dataset; the row for $A>B$ shows the proportion of bootstrap replicates where model A outperformed model B with respect to each metric. Proportions $>0.5$ are highlighted in blue (with darker shades of blue for higher proportions) and proportions $\leq 0.5$ are highlighted in red (with darker shades of red for lower proportions).} 
\label{table:cgn5_fh}
\end{table}

\begin{figure}[!htb]
\centering
\includegraphics[width=0.95\columnwidth]{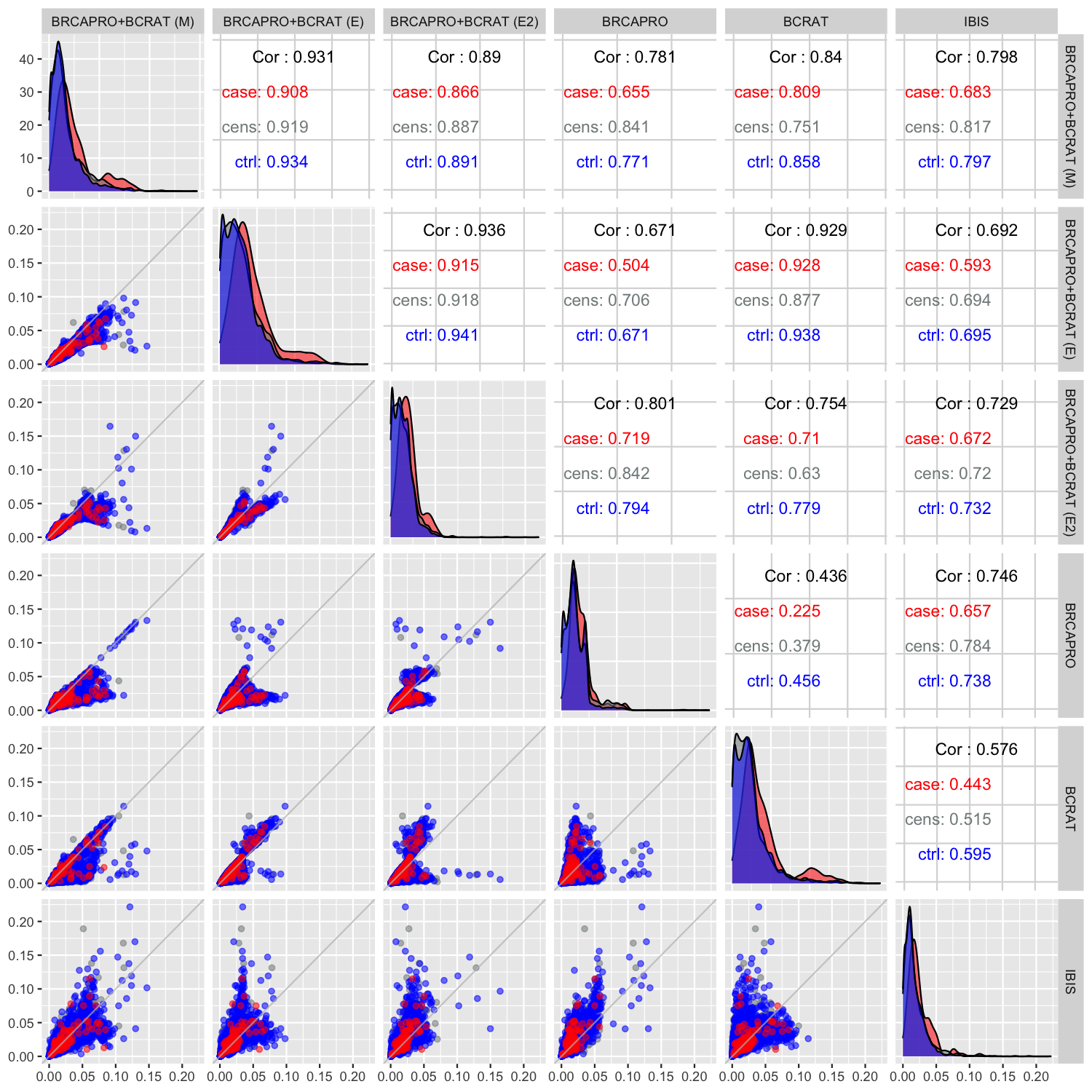}
\caption{Scatter plots, density plots, and correlations stratified by outcome. Red corresponds to cases, blue corresponds to non-cases, and grey corresponds to probands censored before 5 years.}
\label{figure:scatterplots}
\end{figure}

\subsubsection{Time-to-Event Outcome}

The performance measures based on the time-to-event outcome are shown in Tables \ref{table:cgn_fu} (overall performance) and \ref{table:cgn_fu_fh} (performance stratified by family history) in Appendix \ref{appendix:cgn_supp}. BRCAPRO+BCRAT (E) was excluded from the analyses because it only provides 5-year risks. We did not calculate the logarithmic score for BRCAPRO+BCRAT (E2) and IBIS because BRCAPRO+BCRAT (E2) the cause-specific distribution for competing mortality, which is needed to calculate the likelihood, is not explicitly modelled by BRCAPRO+BCRAT (E2) and cannot be easily extracted from the software for IBIS.

The relative performance of the models for the time-to-event outcome was similar to the relative performance for the 5-year outcome. O/E and discriminatory accuracy did not change substantially compared to Tables \ref{table:cgn5} and \ref{table:cgn_fu_fh}, and differences in Brier score and logarithmic score across models were small. BRCAPRO+BCRAT (M) and BRCAPRO+BCRAT (E2)  performed similarly to BCRAT and IBIS overall (Table \ref{table:cgn_fu}) and had higher C-statistics than BCRAT in the subset of probands meeting the NCCN criteria (Table \ref{table:cgn_fu_fh}). The combination models also had better calibration and discrimination than BRCAPRO overall and within each family history stratum.

\section{Discussion and Conclusion}

The availability and use of multiple risk prediction models can lead to confusion in clinical practice. Combining models addresses this problem and also provides a way to develop more comprehensive models without building new models from the ground up, which can save a considerable amount of time and effort. We combined BRCAPRO, a Mendelian model based on detailed family history information, and BCRAT, an empirical model based on a simple summary of family history and various non-genetic risk factors, using two approaches, penetrance modification and ensemble learning. The penetrance modification model, BRCAPRO+BCRAT (M), and the ensemble models, BRCAPRO+BCRAT (E) for binary outcomes and BRCAPRO+BCRAT (E2) for time-to-event outcomes, all achieved accuracy improvements over BRCAPRO and BCRAT in simulations and  data from the CGN, showing the value of augmenting the family history input to BRCAPRO with non-genetic risk factors and augmenting the BCRAT risk factors with more detailed family history information. 

In data simulated under the penetrance modification model, the combination models outperformed BRCAPRO and BCRAT with respect to calibration, discrimination, net benefit, and prediction accuracy based on the Brier score and logarithmic score. In the CGN cohort, where we also validated IBIS, an existing model that combines detailed family history with non-genetic risk factors, the penetrance modification model achieved comparable performance to IBIS, outperforming BRCAPRO overall and achieving better discrimination and net benefit than BCRAT among women with a strong family history of breast/ovarian cancer. The ensemble models also outperformed BRCAPRO overall and outperformed BCRAT among women with a strong family history, but had worse overall calibration than the penetrance modification model. While BCRAT performed well in the entire CGN cohort, the fact that it had less discrimination and net benefit than IBIS and the combination models in probands meeting the NCCN criteria highlights the importance of collecting detailed family history information for higher-risk subgroups where early screening and prevention measures can substantially reduce cancer risk and mortality \cite{metcalfe2008international}. Furthermore, BCRAT is not suitable for known BRCA1/2 carriers, while the other four models all take into account genetic testing results (the ensemble models do so indirectly through the BRCAPRO risk prediction). 

Missing information on BRCAT and IBIS risk factors in the CGN dataset (atypical hyperplasia, age at menarche, and, for IBIS, breast density, hormone replacement therapy, and polygenic risk scores) could potentially have affected the discrimination of BCRAT, IBIS, and the combination models, but they still had relatively good discrimination. The CGN also did not collect genetic testing information for non-probands, which could considerably improve the discrimination of BRCAPRO, IBIS, and the combination models \cite{gail2019performance}. Further validation in independent prospective studies is needed, especially larger studies with more complete covariate information.

One limitation of applying the ensemble models trained on NWH data to the CGN cohort is that the training cohort was not representative of the validation cohort. The NWH cohort was a lower-risk cohort (as seen in Table \ref{data}, it had a lower proportion of women with a first-degree family history of breast cancer) and the family history information available for the NWH cohort was less detailed than that for the CGN cohort. We used importance weighting to address this limitation, but since this approach relies on accurate estimation of the probability distributions of risk factors in the training and target populations, the ensemble models we trained may have been sub-optimal for CGN participants. 
The performance of the ensemble approach could potentially be improved by training on data that is more representative of the validation data. This is supported by fact that the ensemble models were well-calibrated in the simulations, where the training and validation datasets were both generated under BRCAPRO+BCRAT (M).

The two combination approaches each have their strengths and limitations. Ensembling via logistic regression calibrates the model to the training data, which can be a strength or a limitation depending on how well the training data represents the target population. The model might require recalibration in order to be suitable for a population different from the training population. Differences between the training and validation populations could also negatively affect discrimination and other performance measures. The penetrance modification model, on the other hand, does not require training, but relies on accurate estimates of prevalence, penetrance, and relative hazards. These estimates should be updated as new information becomes available. Another disadvantage of ensembling via logistic regression is the need to estimate the censoring distribution when there are probands in the training dataset who are censored before the last time point of interest. One advantage of ensembling is its greater flexibility compared to the penetrance modification approach. Ensembling can easily handle any number of models that can be of any form. Including additional risk factors is also straightforward. The penetrance modification model requires more assumptions because it specifically combines a model based on Mendelian inheritance with a relative hazard model. Additional risk modifiers can be incorporated as new relative hazard estimates become available, but it is important to properly scale the relative hazards to be compatible with the hazard functions they are meant to modify and to consider whether the effects of the risk modifiers differ by carrier status. BRCAPRO+BCRAT (M) currently modifies only the non-carrier hazard function using the BCRAT relative hazard. Future work is needed to extend the model to include modifiers of the carrier hazard functions. One more advantage of ensembling is that once the model is trained, it only requires the predictions from the models being combined and not the raw inputs (besides any additional risk factors that are explicitly included in the ensemble model), which are potentially less accessible than the predictions. 

Given our validation results, the penetrance modification approach seems more promising than ensembling in the context of breast cancer risk prediction. BRCAPRO+BCRAT (M) achieved competitive performance by leveraging the strengths of BRCAPRO and BCRAT, improving on aspects of both models.

\section*{Acknowledgements}
The authors would like to thank Sining Chen and Yufan Liu for their contributions to the formulation of the BRCAPRO+BCRAT (M) model. 

\section*{Code}

The simulation code is available at \url{https://github.com/zoeguan/brcapro_bcrat_combination}.

% Submissions are not required to reflect the precise reference formatting of the journal (use of italics, bold etc.), however it is important that all key elements of each reference are included.
\bibliographystyle{nihunsrt}
\bibliography{absref.bib}

\newpage
\appendix

\section*{Appendices}

\renewcommand{\thesubsection}{\Alph{subsection}}

\subsection{Penetrance Modification Parameters}\label{appendix:penmod_params}

\subsubsection*{Hazard Functions in BRCAPRO and BCRAT}

\begin{figure}[ht]
\centering
\includegraphics[width=0.8\columnwidth]{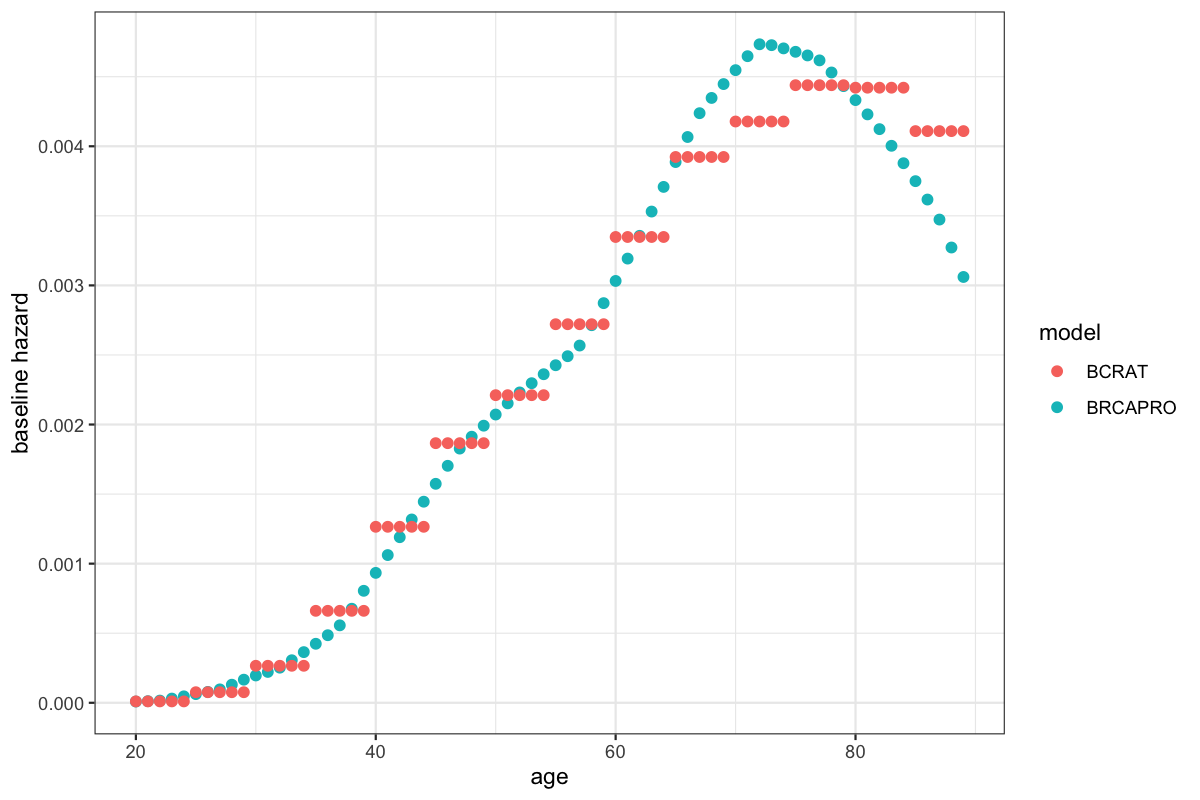}
\caption{BCRAT cause-specific hazard of breast cancer for White women in the general population ($\tilde\lambda_{B}(t) = \tilde\lambda_{B, 0}(t)/(1-AR(t))$) and BRCAPRO cause-specific hazard of breast cancer for White female non-carriers ($\lambda_B^0(t)$).}
\label{figure:hazards}
\end{figure}

\subsubsection*{Population Attributable Risk Estimates for BCRAT Covariates}

\begin{table}[H]
\centering
\begin{tabular}{rrrrrr}
  \hline
 & White & Black & Hispanic & Asian & Native American \\ 
  \hline
$<50$ & 1.81 & 1.41 & 1.37 & 2.10 & 1.55 \\ 
  $\geq 50$ & 1.96 & 1.44 & 1.41 & 2.43 & 1.94 \\ 
   \hline
\end{tabular}
\caption{Estimates of $(1-AR(t))$ from NHIS 2015.}
\end{table}

\newpage
\subsection{Ensemble Weights From NWH}\label{appendix:weights}

For BRCAPRO + BCRAT (E), we fit the model
\begin{equation*}
\log \frac{F_B(\tau)}{1-F_B(\tau)} = \beta_0 + \beta_1 \sqrt{F_B^{1}(\tau)} + \beta_2 \sqrt{F_B^{2}(\tau)} + \beta_3 \sqrt{F_B^{1}(\tau)} \sqrt{F_B^{2}(\tau)},
\end{equation*}
to the NWH cohort, where $\tau=5$. The estimated weights are given below.

\begin{table}[ht]
\centering
\begin{tabular}{lll}
  \hline
 & Estimate & Standard Error \\ 
  \hline
$\hat \beta_0$ & 2.55 & 1.31 \\ 
$\hat \beta_1$ & 0.86 & 0.32 \\ 
$\hat \beta_2$ & 1.21 & 0.28 \\ 
$\hat \beta_3$ & 0.11 & 0.06 \\ 
   \hline
\end{tabular}
\caption{Coefficient estimates for BRCAPRO + BCRAT (E).}
\label{table:ensemble_weights}
\end{table}

For BRCAPRO + BCRAT (E2), we fit the model
\begin{flalign*}
\log \frac{F_B(\tau)}{1-F_B(\tau)} &= \beta_0 + \beta_1 \sqrt{F_B^{1}(\tau)} + \beta_2 \sqrt{F_B^{2}(\tau)} + \beta_3 \sqrt{F_B^{1}(\tau)} \sqrt{F_B^{2}(\tau)} +
\beta_4 \tau +
\beta_5 \tau \sqrt{F_B^{1}(\tau)} + \beta_6 \tau \sqrt{F_B^{2}(\tau)} +  \\
&\phantom{==} \beta_7 \tau \sqrt{F_B^{1}(\tau)}\sqrt{F_B^{2}(\tau)}.
\end{flalign*}
to the NWH cohort using observations from years $\tau=1,\dots,8$. The estimated weights are given below.

\begin{table}[ht]
\centering
\begin{tabular}{rrr}
  \hline
 & Estimate & Standard Error \\ 
  \hline
$\hat \beta\_0$ & -10.28 & 0.99 \\ 
  $\hat \beta\_1$ & 43.11 & 11.53 \\ 
  $\hat \beta\_2$ & 38.30 & 7.67 \\ 
  $\hat \beta\_3$ & 0.35 & 0.13 \\ 
  $\hat \beta\_4$ & -257.23 & 79.50 \\ 
  $\hat \beta\_5$ & -3.21 & 1.42 \\ 
  $\hat \beta\_6$ & -2.42 & 0.92 \\ 
  $\hat \beta\_7$ & 22.98 & 9.80 \\ 
   \hline
\end{tabular}
\caption{Coefficient estimates for BRCAPRO + BCRAT (E2).}
\label{table:ensemble_weights_2}
\end{table}

\newpage
\subsection{CGN Validation - Additional Tables}\label{appendix:cgn_supp}

\begin{table}[ht]
\footnotesize
\centering
\begin{tabular}{lllllllll}
  \hline
Variable & N & Age &  \multicolumn{3}{l}{Affected 1st-degree Relatives} & Follow-up & Censored & Cases \\
 &  & \scalebox{.85}[1.0]{(median [IQR])} & (\%) & & & \scalebox{.85}[1.0]{(median [IQR])} & (\%) & (\%) \\
  Category &  &  & 0 & 1 & 2+ &  &  & \\
      \hline
  \scalebox{.85}[1.0]{BAYLOR} & 69 & 47 [38, 52] & 34 (49.3) & 28 (40.6) & 7 (10.1) & 7.5 [6.1, 8.6] & 12 (17.4) & 0 (0.0) \\ 
  \scalebox{.85}[1.0]{COLORADO} & 1198 & 51 [41, 64] & 528 (44.1) & 547 (45.7) & 123 (10.3) & 7.6 [6.4, 8.5] & 85 (7.1) & 23 (1.9) \\ 
  \scalebox{.85}[1.0]{DUKE} & 286 & 46 [38, 53] & 134 (46.9) & 116 (40.6) & 36 (12.6) & 7.2 [6.2, 8.2] & 40 (14.0) & 9 (3.1) \\ 
  \scalebox{.85}[1.0]{EMORY} & 136 & 44 [38.8, 51] & 56 (41.2) & 51 (37.5) & 29 (21.3) & 7.2 [6.6, 8.3] & 29 (21.3) & 3 (2.2) \\ 
  \scalebox{.85}[1.0]{GEORGETOWN} & 309 & 43 [35, 52] & 107 (34.6) & 147 (47.6) & 55 (17.8) & 7.8 [6.3, 8.5] & 81 (26.2) & 2 (0.6) \\ 
  \scalebox{.85}[1.0]{JH} & 469 & 47 [39, 56] & 279 (59.5) & 148 (31.6) & 42 (9.0) & 8.0 [6.1, 9.0] & 73 (15.6) & 10 (2.1) \\ 
  \scalebox{.85}[1.0]{MDAND} & 295 & 45 [37, 53] & 215 (72.9) & 64 (21.7) & 16 (5.4) & 6.1 [4.2, 7.0] & 87 (29.5) & 4 (1.4) \\ 
  \scalebox{.85}[1.0]{UCI} & 608 & 48 [37, 59] & 352 (57.9) & 223 (36.7) & 33 (5.4) & 5.3 [4.0, 7.1] & 209 (34.4) & 4 (0.7) \\ 
  \scalebox{.85}[1.0]{UNC} & 229 & 46 [39, 53] & 80 (34.9) & 111 (48.5) & 38 (16.6) & 8.0 [7.1, 9.0] & 28 (12.2) & 6 (2.6) \\ 
  \scalebox{.85}[1.0]{UNM} & 324 &  [41, 63] & 160 (49.4) & 123 (38.0) & 41 (12.7) & 6.6 [6.0, 7.5] & 43 (13.3) & 11 (3.4) \\ 
  \scalebox{.85}[1.0]{UPENN} & 540 & 45 [37, 53] & 297 (55.0) & 185 (34.3) & 58 (10.7) & 8.1 [6.6, 9.1] & 76 (14.1) & 8 (1.5) \\ 
  \scalebox{.85}[1.0]{UTAH} & 880 & 47 [35, 61] & 522 (59.3) & 298 (33.9) & 60 (6.8) & 7.3 [5.4, 8.0] & 61 (6.9) & 14 (1.6) \\ 
  \scalebox{.85}[1.0]{UTSA} & 92 & 43 [35.8, 52] & 29 (31.5) & 48 (52.2) & 15 (16.3) & 5.1 [4.0, 6.5] & 36 (39.1) & 1 (1.1) \\ 
  \scalebox{.85}[1.0]{UTSW} & 247 & 41 [33.5, 47] & 88 (35.6) & 116 (47.0) & 43 (17.4) & 6.6 [5.6, 7.6] & 30 (12.1) & 4 (1.6) \\ 
  \scalebox{.85}[1.0]{UWASH} & 1632 & 46 [37, 56] & 1290 (79.0) & 291 (17.8) & 51 (3.1) & 7.6 [6.9, 8.3] & 44 (2.7) & 13 (0.8) \\ 
   \hline
\end{tabular}
\caption{CGN cohort characteristics by center.} \label{table:cgn_centers}
\end{table}

\begin{table}[!htbp]
\centering
\footnotesize
\begin{tabular}{lllll}
  \hline
 & O/E & C-statistic & $\Delta$BS & $\Delta$LS \\ 
  \hline
\textbf{Performance Metrics} & & & & \\
B+B (M) & 0.98 (0.84, 1.13) & 0.68 (0.64, 0.71) & 0.25 (-0.50, 0.95) & 0.45 (-0.13, 0.96) \\ 
  B+B (E) & 1.08 (0.92, 1.23) & 0.68 (0.63, 0.72) & 0.35 (-0.22, 0.86) &  \\ 
  BRCAPRO & 1.25 (1.07, 1.43) & 0.64 (0.60, 0.68) & 0.00 (0.00, 0.00) & 0.00 (0.00, 0.00) \\ 
  BCRAT & 1.08 (0.92, 1.24) & 0.67 (0.63, 0.71) & 0.32 (-0.55, 1.12) & -0.76 (-1.64, 0.11) \\ 
  IBIS & 0.93 (0.79, 1.07) & 0.67 (0.63, 0.70) & -0.08 (-0.88, 0.64) &  \\ 
\multicolumn{3}{l}{\textbf{Comparisons Across Bootstrap Replicates}} & & \\
B+B(M)>B+B(E2) & \mycolor{0.641} & \mycolor{0.421} & \mycolor{0.331} &  \\ 
  B+B(M)>BRCAPRO & \mycolor{0.903} & \mycolor{0.994} & \mycolor{0.735} & \mycolor{0.943} \\ 
  B+B(M)>BCRAT & \mycolor{0.648} & \mycolor{0.642} & \mycolor{0.383} & \mycolor{0.988} \\ 
  B+B(M)>IBIS & \mycolor{0.707} & \mycolor{0.749} & \mycolor{0.822} &  \\ 
  B+B(E2)>BRCAPRO & \mycolor{0.970} & \mycolor{0.998} & \mycolor{0.887} &  \\ 
  B+B(E2)>BCRAT & \mycolor{0.579} & \mycolor{0.710} & \mycolor{0.569} &  \\ 
  B+B(E2)>IBIS & \mycolor{0.476} & \mycolor{0.737} & \mycolor{0.873} &  \\ 
   \hline
\end{tabular}
\caption{Performance over follow-up period in entire CGN cohort. B+B: BRCAPRO+BCRAT. $\Delta$BS: \% relative improvement in Brier Score compared to BRCAPRO. $\Delta$LS: \% relative improvement in logarithmic score compared to BRCAPRO. The ``Comparisons Across Bootstrap Replicates" section shows pairwise comparisons involving the combination models across 1000 bootstrap replicates of the validation dataset; the row for $A>B$ shows the proportion of bootstrap replicates where model A outperformed model B with respect to each metric. Proportions $>0.5$ are highlighted in blue (with darker shades of blue for higher proportions) and proportions $\leq 0.5$ are highlighted in red (with darker shades of red for lower proportions).}
\label{table:cgn_fu}
\end{table}

\begin{table}[!htbp]
\centering
\footnotesize
\begin{tabular}{lllll}
  \hline
 & O/E & C-statistic & $\Delta$BS & $\Delta$LS \\ 
  \hline
   \multicolumn{3}{l}{\underline{\textbf{Strong Family History (45 cases)}}} & & \\
\textbf{Performance Metrics} & & & & \\
B+B (M) & 0.85 (0.62, 1.09) & 0.67 (0.59, 0.75) & -0.04 (-2.00, 1.54) & 0.07 (-2.63, 2.15) \\ 
  B+B (E) & 1.10 (0.81, 1.41) & 0.65 (0.57, 0.73) & -0.10 (-1.37, 1.16) &  \\ 
  BRCAPRO & 1.40 (1.03, 1.81) & 0.63 (0.54, 0.72) & 0.00 (0.00, 0.00) & 0.00 (0.00, 0.00) \\ 
  BCRAT & 1.05 (0.77, 1.35) & 0.61 (0.52, 0.70) & -0.12 (-2.48, 2.00) & -3.36 (-7.07, -0.29) \\ 
  IBIS & 0.77 (0.56, 0.98) & 0.68 (0.59, 0.78) & -0.75 (-3.14, 0.99) &  \\  
\multicolumn{3}{l}{\textbf{Comparisons Across Bootstrap Replicates}} & & \\
B+B(M)>B+B(E2) & \mycolor{0.385} & \mycolor{0.930} & \mycolor{0.522} &  \\ 
  B+B(M)>BRCAPRO & \mycolor{0.781} & \mycolor{0.916} & \mycolor{0.464} & \mycolor{0.501} \\ 
  B+B(M)>BCRAT & \mycolor{0.322} & \mycolor{0.963} & \mycolor{0.525} & \mycolor{0.943} \\ 
  B+B(M)>IBIS & \mycolor{0.954} & \mycolor{0.321} & \mycolor{0.808} &  \\ 
  B+B(E2)>BRCAPRO & \mycolor{0.933} & \mycolor{0.685} & \mycolor{0.420} &  \\ 
  B+B(E2)>BCRAT & \mycolor{0.336} & \mycolor{0.953} & \mycolor{0.534} &  \\ 
  B+B(E2)>IBIS & \mycolor{0.717} & \mycolor{0.144} & \mycolor{0.770} &  \\ 
  \hline
\multicolumn{3}{l}{\underline{\textbf{Less Family History (114 cases)}}} & & \\
\textbf{Performance Metrics} & & & & \\
B+B (M) & 1.18 (0.97, 1.39) & 0.68 (0.63, 0.73) & 0.54 (-0.06, 1.11) & 0.53 (0.07, 1.06) \\ 
  B+B (E) & 1.17 (0.96, 1.38) & 0.68 (0.64, 0.72) & 0.59 (0.14, 1.07) &  \\ 
  BRCAPRO & 1.31 (1.08, 1.55) & 0.63 (0.58, 0.68) & 0.00 (0.00, 0.00) & 0.00 (0.00, 0.00) \\ 
  BCRAT & 1.20 (0.99, 1.42) & 0.69 (0.65, 0.73) & 0.65 (0.06, 1.26) & -0.22 (-1.09, 0.50) \\ 
  IBIS & 1.14 (0.94, 1.35) & 0.68 (0.64, 0.72) & 0.31 (-0.22, 0.82) &  \\ 
\multicolumn{3}{l}{\textbf{Comparisons Across Bootstrap Replicates}} & & \\
B+B(M)>B+B(E2) & \mycolor{0.153} & \mycolor{0.432} & \mycolor{0.361} &  \\ 
  B+B(M)>BRCAPRO & \mycolor{0.990} & \mycolor{0.991} & \mycolor{0.954} & \mycolor{0.992} \\ 
  B+B(M)>BCRAT & \mycolor{0.966} & \mycolor{0.115} & \mycolor{0.230} & \mycolor{0.958} \\ 
  B+B(M)>IBIS & \mycolor{0.060} & \mycolor{0.593} & \mycolor{0.741} &  \\ 
  B+B(E2)>BRCAPRO & \mycolor{0.985} & \mycolor{0.999} & \mycolor{0.996} &  \\ 
  B+B(E2)>BCRAT & \mycolor{0.964} & \mycolor{0.163} & \mycolor{0.281} &  \\ 
  B+B(E2)>IBIS & \mycolor{0.062} & \mycolor{0.603} & \mycolor{0.844} &  \\ 
   \hline
\end{tabular}
\caption{Performance over follow-up period in CGN cohort, stratified by family history (whether or not the proband met the NCCN criteria for further genetic risk evaluation \cite{nccn}; in applying the criteria, we only used information on breast and ovarian cancer diagnoses in relatives). B+B: BRCAPRO+BCRAT. $\Delta$BS: \% relative improvement in Brier Score compared to BRCAPRO. The ``Comparisons Across Bootstrap Replicates" section shows pairwise comparisons involving the combination models across 1000 bootstrap replicates of the validation dataset; the row for $A>B$ shows the proportion of bootstrap replicates where model A outperformed model B with respect to each metric. Proportions $>0.5$ are highlighted in blue (with darker shades of blue for higher proportions) and proportions $\leq 0.5$ are highlighted in red (with darker shades of red for lower proportions).}
\label{table:cgn_fu_fh}
\end{table}

\end{document}